\documentclass[
aps,
reprint,
superscriptaddress,
amsmath,
amssymb
]{revtex4-1}

\usepackage{times}
\usepackage{hyperref}
\usepackage[T1]{fontenc}
\usepackage[german,ngerman,english]{babel}

\usepackage{mathtools} 
\usepackage{amsmath} 
\usepackage{bbm,bm}
\usepackage{dsfont}
\usepackage{braket}
\usepackage{amsthm}
\usepackage{enumerate}
\usepackage{enumitem}
\usepackage{float}
\usepackage{ifthen}
\usepackage[normalem]{ulem}

\usepackage{tikz}
\usepackage{tikzscale}
\usetikzlibrary{calc}
\usetikzlibrary{decorations.pathreplacing}
\usetikzlibrary{decorations}
\usetikzlibrary{positioning}
\usetikzlibrary{plotmarks}
\usetikzlibrary{arrows.meta}
\usepackage{pgfplots}
\usepgfplotslibrary{patchplots}
\pgfplotsset{compat=newest}
\usepackage{xxcolor}
\definecolor{structure}{rgb}{0.23,0.4,0.7}

\newtheorem{theorem}{Theorem}
\newtheorem{lemma}[theorem]{Lemma}

\newtheorem{definition}[theorem]{Definition}

 \newtheorem{proposition}[theorem]{Proposition}

 \newcommand{\e}{\mathrm{e}}
 \renewcommand{\i}{\mathrm{i}}

 \newcommand{\CNOT}{\mathrm{CNOT}}
  \newcommand{\Tr}{{\rm Tr}}

 \newcommand{\EE}{\mathbb{E}}
 \newcommand{\E}{\mathbb{E}}
 \newcommand{\1}{\mathds{1}}
 \newcommand{\mb}[1]{\mathbb{#1}}
 
 \newcommand{\Z}{\mb{Z}}
 
\newcommand{\mc}[1]{\mathcal{#1}}

\newcommand{\mcB}{\mc{B}}

\newcommand{\mcD}{\mc{D}}
\newcommand{\mcG}{\mc{G}}
\newcommand{\mcC}{\mc{C}}
\newcommand{\mcU}{\mc{U}}

\newcommand{\coloneqq}{\mathrel{\vcentcolon\mkern-1.2mu=}} 

\renewcommand{\max}{\mathchoice{\operatorname*{max}}{\operatorname*{max}}{\mathrm{max}}{\mathrm{max}}} 

\newcommand{\norm}[2][]{
  \ifthenelse{\equal{#1}{}}
    {\left\| {#2} \right\|}
    {\ifthenelse{\equal{#1}{uinv}}
      {\left\vert\kern-0.25ex\left\vert\kern-0.25ex\left\vert {#2} \right\vert\kern-0.25ex\right\vert\kern-0.25ex\right\vert}
      {\left\| {#2} \right\|_{#1}}
    }
}

\newcommand{\taverage}[2][]{
  \ifthenelse{\equal{#1}{}}
  {\overline{#2}}
  {\overline{#2}^{#1}}
}

\newcommand{\tracedistance}[3][]{
  \ifthenelse{\equal{#2}{}}
  {\ifthenelse{\equal{#3}{}}
    {\mathcal{D}_{#1}}{}
  }{
    \ifthenelse{\equal{#1}{}}
    {\mathchoice{\operatorname{\mathcal{D}}\left(#2,#3\right)}{\operatorname{\mathcal{D}}(#2,#3)}{\operatorname{\mathcal{D}}(#2,#3)}{\operatorname{\mathcal{D}}(#2,#3)}}
    {\mathchoice{\operatorname{\mathcal{D}}_{#1}\left(#2,#3\right)}{\operatorname{\mathcal{D}}_{#1}(#2,#3)}{\operatorname{\mathcal{D}}_{#1}(#2,#3)}{\operatorname{\mathcal{D}}_{#1}(#2,#3)}}
  }
}
\newcommand{\fidelity}[3][]{
  \ifthenelse{\equal{#2}{}}
  {\ifthenelse{\equal{#3}{}}
    {\mathcal{F}_{#1}}{}
  }{
    \ifthenelse{\equal{#1}{}}
    {\mathchoice{\operatorname{\mathcal{F}}\left(#2,#3\right)}{\operatorname{\mathcal{F}}(#2,#3)}{\operatorname{\mathcal{F}}(#2,#3)}{\operatorname{\mathcal{F}}(#2,#3)}}
    {\mathchoice{\operatorname{\mathcal{F}}_{#1}\left(#2,#3\right)}{\operatorname{\mathcal{F}}_{#1}(#2,#3)}{\operatorname{\mathcal{F}}_{#1}(#2,#3)}{\operatorname{\mathcal{F}}_{#1}(#2,#3)}}
  }
}

\newcommand{\Sr}[3][]{
  \ifthenelse{\equal{#1}{}}
    {\operatorname{\mathnormal{S}}(#2\|#3)}
    {\operatorname{\mathnormal{S}}_{#1}(#2\|#3)}
}



\newcommand{\je}[1]{{\color{black} #1}}
\definecolor{eo}{RGB}{246,76,25}
\newcommand{\eo}[1]{{\color{black} #1}}

\definecolor{eo}{RGB}{246,76,25}

\begin{document}

\title{Randomized benchmarking for individual quantum gates}
\author{E. Onorati},
\address{Dahlem Center for Complex Quantum Systems, Freie Universit\"{a}t Berlin, 14195 Berlin, Germany}

\author{A. H. Werner}
\address{Department of Mathematical Sciences, University of Copenhagen, DK-2100 K{\o}benhavn, Denmark}
\author{J. Eisert}
\address{Dahlem Center for Complex Quantum Systems, Freie Universit\"{a}t Berlin, 14195 Berlin, Germany}
\address{Department of Mathematics and Computer Science, Freie Universit\"{a}t Berlin, 14195 Berlin, Germany}

\begin{abstract}
Any technology requires precise benchmarking of its components, and the quantum technologies are no exception. Randomized benchmarking allows for the relatively resource economical estimation of the average gate fidelity of quantum gates from the Clifford group, assuming identical noise levels for all gates, making use of suitable sequences of randomly chosen Clifford gates. In this work, we report significant progress on randomized benchmarking, by showing that it can be done for individual quantum gates outside the Clifford group, even for varying noise levels per quantum gate. This is possible at little overhead of quantum resources, but at the expense of a significant classical computational cost. At the heart of our analysis is a representation-theoretic framework that we develop here which is brought into contact with classical estimation techniques based on bootstrapping and matrix pencils. We demonstrate the functioning of the scheme at hand of benchmarking tensor powers of $T$-gates. Apart from its practical relevance, we expect this insight to be relevant as it highlights the role of assumptions made on unknown noise processes when characterizing quantum gates at high precision.
\end{abstract}

\maketitle

 Recent years have seen a rapid development in the precise preparation and manipulation of quantum states. Much of these advances
 have been triggered by the progress in the quantum technologies, anticipating computational quantum devices that outperform
 classical computers \cite{Roadmap}.
 The increased sophistication of these experiments, however, is accompanied by an increased complexity of
 the task of verifying that the process has actually been implemented as expected. Unwanted and usually largely
 uncharacterized quantum noise hampers the correct functioning of quantum devices. Without the knowledge of
such noise processes, yet, it is unclear how to achieve actionable advice on how to improve the experiment
or whether the experiment has worked at all. The task of certifying quantum processes has, unsurprisingly, hence been
identified as a key bottleneck in the field.
	
There are several strategies that can in principle be employed to tackle the problem. They can largely be classified in the assumptions they
make about the given unknown process and the structure they employ. \emph{Quantum process tomography} \cite{ChuangNielsenProcessTomography,PhysRevA.77.032322} is an option, in principle, in that
based on measurement data alone, an unknown process can be reconstructed. However, without any further structure or assumptions the scaling of
the effort of quantum process tomography is daunting: for a generic $d$-dimensional quantum system, $\Theta(d^4)$ expectation
values must be estimated to great accuracy. Methods exploiting structure, such as process tomography
based on compressed sensing \cite{Compressed,AverageGateFidelities} or diamond norm minimization \cite{Diamond}
reduce the effort significantly. Also, the dynamics of quantum systems for short times
can often be kept track of using tensor network states \cite{QuantumFieldTomography,MPSTomographyIons}. Nevertheless, the benchmarking of large quantum circuits
involving a large number of gates is out of scope in either of these {approaches}. This challenge is aggravated by the fact that
gate errors {suitable for \emph{fault tolerant quantum computing}}
\cite{PreskillFaultTolerant,MartinisFowlerFaultTolerant,IBMFaultTolerant,Roadmap} are extremely small, so that the characterization
needs to be very precise.

\emph{Randomized benchmarking} 
\je{\cite{FirstRB,PhysRevA.80.012304,KnillBenchmarking,MagGamEmer,MagGamEmer2,RBFlammia,EmersonDiscussion} -- 
{introduced in Ref.\ \cite{FirstRB} --}}
takes a different, more pragmatic route altogether.  It acknowledges that full tomographic
knowledge may be too much to ask for and achieves estimates of quantum gate errors making use of long sequences of randomly chosen Clifford gates. This is done to estimate a single quantity, the \emph{average gate fidelity}, for a set of operations being the \emph{Clifford group} in a large portion of the literature.
That is to say, by making stronger assumptions about the underlying operations, one can go a long
way reliably characterizing noise levels of quantum gates. The basic idea has been
generalised in several ways;
in particular, it
has been shown that one does not
need to employ an exact so-called \emph{2-design} for randomized benchmarking~\je{\cite{PhysRevA.80.012304,GroAudEis,Cross16}}, 
using instead, for instance, the single-qubit dihedral group~\cite{DugWallEme15}.
To allow for extraction of the fidelity of a certain gate, a scheme called \emph{interleaved randomized benchmarking} \cite{InterleavedRB,Harper} makes use of random sequences of Clifford gates interlaced by the particular gate whose noise is to be individually characterized, but it is still limited to the Clifford group itself or the $T$-gate. 
\eo{For protocols based on twirls over 2-design gates sets, such as the Clifford group, it has been recently shown~\cite{IndependentNoise} that they allow for arbitrarily large and gate-dependent noise levels. However, other schemes which are able to go beyond the 2-design requirement assume 
the noise to be gate-independent~\cite{RBFiniteGroups}, or weakly dependent~\cite{WintonRB}.}
\eo{In several ways the commonly made assumptions on the underlying processes are rather strong. For protocols not relying on 2-designs property, the premise on uniform noise channel for each gate of the group may be overly demanding; even if we take into consideration small generating gates sets, it is difficult to postulate that the noise of a single gate is equal to the one of a product of several consecutive operators.} Here, we will see that some of these assumptions can be
 relaxed at little additional cost, as far as quantum resources are concerned.

In this work, we introduce the notion of randomized benchmarking for individual quantum
gates, one that may be part of single layers of gate arrays.
This work is expected to be significant in two ways: technically speaking, we introduce methods of randomized benchmarking able to
benchmark individual quantum gates, including ones that are outside the Clifford group. Since schemes for universal quantum computing necessarily make use
of such gates, this seems an important step forward. At the same time, we do not require twirling over the full Clifford group or a 2-design, but only over a relatively small local symmetry group coupled with transpositions gates. \je{It is a favourable feature of our approach to
make use of a small group in contrast to large gate sets of previously known schemes.}

The novel idea in the present work is to exploit the symmetry of the
quantum gate itself in an appropriate fashion and hence reduce the amount of quantum resources and computational effort.
To achieve this goal, we harness advanced tools from representation theory to
arrive at schemes that require similar physical operations, departing from the paradigm of
2-designs, but which can
make predictions beyond known prescriptions. 
{These ideas are uplifted to functioning schemes by making use of \emph{estimation techniques}
such as matrix pencils \cite{MatrixPencilMethodDumped}. It is key to the new approach that the demands for classical estimation are higher, which we
accommodate by introducing more sophisticated tools to this task. Using such methods, we demonstrate that meaningful and experimentally
accessible preparations are sufficient to render the recovery feasible at hand of randomized benchmarking data.}

More conceptually speaking, and putting this contribution into a broader context,
we show that one can interpolate between common assumptions made when
characterizing quantum processes: In other words, there is ``room in the middle''
between full quantum process tomography, which is largely assumption-free but
comes along with daunting resource requirements, and conventional randomized benchmarking,
which requires significantly less effort and is also robust against \emph{state preparation and measurement (SPAM)}
errors, {while making}
strong assumptions. {Complementing this mindset in a very different way,
Refs.\ \cite{AverageGateFidelities,KimmOhki} provide method to 
extract tomographic information making use standard randomized benchmarking protocols, while 
maintaining very large data levels.}
We believe
that this conceptual insight into the ontology of assumptions when characterizing
quantum processes subject to 
{quantum noise} is equally important.

\paragraph*{{Exploiting local symmetries in randomized benchmarking.}}
In the following, we are going to describe a protocol that provides the \emph{average gate fidelity} of the noisy
channel $\Lambda$ which characterizes the imperfect implementation  $\widetilde U$
of a target ideal unitary gate $U$,
\begin{equation}
	\EE({\mathcal F _{\widetilde \mcU, \mcU}})= \EE({\mathcal F_{\Lambda, \mathcal I}})
	\coloneqq
	\int_{\mathrm{Haar}} \Tr\left[ \ket{\phi}\bra{\phi} \, \Lambda (\ket{\phi}\bra{\phi}) \right] \, \mathrm d \phi.
\end{equation}
It is key to our method to explicitly exploit the local and permutation symmetries of $U$,
allowing for a drastic reduction of the fitting parameters and  also inherits robustness with respect to SPAM errors.
In this way, fitting models of well-known randomized benchmarking protocols can be uplifted to this setting
involving fewer assumptions.

Throughout this work, we consider quantum gates acting on $n$-qubit systems and we are interested in benchmarking the accuracy of their implementation in a quantum circuit making use of the protocol that we are going to explain in a later section. The method is particularly suitable for gates consisting of tensor products of local gates, hence admitting additional symmetries with respect to the  exchange of qubit subsystems and hence it is applicable in those
situations of single layers of local unitary quantum gates.
For concreteness,
we shall
put emphasis on single layers of circuits whose gates consist of tensor compositions of the $T$-gate with {other} 
gates  belonging to the Clifford group $\mathcal C_n$, namely $H$, {$S$} and CNOT
\begin{align}\label{eq:set_universal_gates}
H&=\frac{1}{\sqrt{2}} \begin{pmatrix}1 &1 \\1 &-1 \end{pmatrix}, \,
&T&= \begin{pmatrix}\e^{-\i \pi/8} &0 \\0 & \e^{\i \pi/8}\end{pmatrix}, \nonumber\\
 \CNOT &= \begin{pmatrix}1&0&0&0\\0&1&0&0\\0&0&0&1\\0&0&1&0\end{pmatrix}, \,&{S}&=\begin{pmatrix}1 & 0 \\ 0 & \i \end{pmatrix},
\end{align}
leading to a universal set. {This setting is of paramount importance in state-of-the-art prescriptions of 
fault-tolerant quantum computing \cite{PreskillFaultTolerant,MartinisFowlerFaultTolerant,IBMFaultTolerant,Roadmap}.}
More generally, the protocol is suitable to benchmark tensor products of local gates  consisting in arbitrary rotations around $X,Y,Z$-axes of the Bloch sphere.
{We} will denote with $U= U_1 \otimes \dots \otimes U_{m}$
the multi-qubit unitary operator -- with $U_j$ {one or two qubits each} -- that we intend
to benchmark, acting as a single layer of a unitary circuit. At the heart of the analysis will be its
symmetry group, constructed from the symmetries of the local gates $U_j$ composing $U$ and the permutations of qubits which the local gate acts upon. More precisely, we choose  the local symmetry group $A_j$ of $U_j$ as the subset of the single-qubit Clifford group whose elements commute with $U_j$. 
{E.g.,}
for $U_j=T= \exp (- \i \pi   Z/8)$, the group of local symmetries {is} 
\begin{equation}
A_T \coloneqq \set{ U \in \mathcal C_1 : U^\dagger \, Z \, U = Z }.
\end{equation}
This is an abelian group of 4 elements isomorphic to the cyclic group of order 4, $\Z_4$.
The set of all possible permutations interchanging qubits affected by the same local gate is another symmetry group of the target unitary $U$; taking a pratical example, for the gate $U=T \otimes H \otimes T \otimes H \otimes T$, this group is isomorphic to $S_3 \times S_2$, i.e., all permutations of the first, third and fifth subsystems combined with the transposition of the second and forth subsystems.  The full symmetry group $G$ is then obtained through the \emph{semi-direct product} $A_n  \rtimes \Pi$, where $A_n$ is the \emph{direct product} of the local symmetry groups $A_j$ constructed by the Kronecker product of the respective elements, and $\Pi$ is the representation of the subgroup of $S_n$ consisting of all allowed permutations of the qubits subsystems.

\paragraph*{{Role of abelian groups}.} In order to apply our full protocol and combine the group $A_n$ with $\Pi$, all local symmetry groups $A_j$ must be abelian. This is indeed a necessary condition to reconstruct the irreducible representations of the full group $G$ with the sole knowledge of the composing groups, as we will discuss in the appendix. Fortunately, this is the case for the symmetry of the gates in Eq.\ \eqref{eq:set_universal_gates} and all other rotations around Bloch axes. Should the local symmetry groups not be all abelian, the protocol is still valid setting $G=A_n$, i.e., without considering permutation symmetries.

\paragraph*{Assumptions and physical motivation.} We  denote with calligraphic letters the channel acting by gate conjugation on density operators, i.e., $\mathcal{U}(\rho) := U^\dagger \rho\, U$ and the noisy implementation of the idealized gate channel  $\mcU$  as $\widetilde \mcU \coloneqq \Lambda_\mcU  \circ \mcU$, i.e., we account for a gate-dependent error channel $\Lambda_\mcU$ whose average fidelity we want to characterize with the proposed protocol.
As randomized benchmarking can be interpreted as a trade-off between the level of characterization of the noise channel and the amount of physical and computational resources needed, we will make the following assumption: the
Pauli-Liouville representation of the twirled noise channel, $\Lambda_\mcU^G \coloneqq {|G|^{-1}} \sum_{j \in \mathbb N_{|G|}} \mcG_j^\dagger \Lambda_\mcU \, \mcG_j $, is \emph{almost jointly diagonalizable} with the target unitary channel $\mcU$ (where again we consider it as a matrix in
Pauli-Liouville representation), in the spirit of Ref.~\cite{GlasBron}. This means that when the matrix $\mathbb{\mcU}$ is brought to diagonal form by some unitary transformation $V$, the off-diagonal element of $\Lambda_\mcU^G$ under the same transformation are small and in particular, $\Lambda_\mcU^G$ will approximately leave the same subspaces invariant. This is true in two cases.
The first possibility is that both $\mcU$ and the twirled noise $\Lambda_\mcU^G$ are diagonalizable simultaneously in a certain basis, e.g. when the decomposition of the representation of the symmetry group into irreducible representations has no multiplicity: in this case, both $\mcU$ and the twirled noise $\Lambda_\mcU^G$ are ``forced'' to assume a diagonal form with respect to the irreducible subspaces.
If this is not the case, then $\Lambda_\mcU^G$ assumes a sparse form with some off-diagonal entries, which have to be small with respect to the diagonal elements.  This is fulfilled whenever the original noise channel $\Lambda_\mcU$ related to the implementation of the gate $\mcU$ was  almost jointly diagonalizable to begin with, or put in another perspective 
{\cite{GlasBron}}, it is almost commuting. This assumption is valid when the gate $U$ is generated by a Hamiltonian $H$ applied for some run time $t$ \cite{AndSosaRioDeuJes}, i.e., $U=\e^{-\i H t}$, which can be perturbed for a small fraction of the time, or be applied for too much or too little time (please refer to Section~\ref{sec:Zassenhaus} in the appendix for more details). Furthermore, we ask the gates belonging to the symmetry group $G$  to be implementable with high accuracy. These gates either perform a permutation of the subspaces of the system or belong to the Clifford group and so can be for instance benchmarked with the well-known protocols~{\cite{MagGamEmer,MagGamEmer2,IndependentNoise}}. 

\paragraph*{The protocol.} We propose a slightly modified version of the previous protocols. We apply in succession channels defined by the gate $U$ after the one induced by a gate uniformly  drawn at random from the symmetry group $G$. In addition, we can target the different symmetry subspaces that are stabilized by $\mathcal{G}$ by choosing an appropriate initial state through the application of projectors decomposing a density operator into basis vectors of distinct irreducible subspaces (cfr. eq.~\eqref{eq:projectors_onto_irred_subspace} in the appendix). Note that, unlike previous protocols, the target gate $U$ is not part of the twirling group $G$: This is one of the reason why one can benchmark arbitrarily small rotations over the Bloch axes with a relatively small number of gates.
For a fixed sequence length $\ell$ the protocol consists of the 
steps:
\begin{enumerate}[label={(\roman*)}]
\item Prepare an initial state $\rho$ with support in the target invariant subspace(s).
\item  Draw a random sequence $\mathbf{k}_{ \ell} =(k_1,\dots, k_\ell) \in \mathbb N_{|G |}^\ell$.
\item Apply the following operation generated by the symmetry operations $\mathcal{G}_{k_i}$ to the initial state $\rho$
	\begin{equation}
	\mathcal C_{\mathbf{k}_\ell}(\rho)  =\mathcal{G}_{\mathrm{inv}} \circ \mathcal{U} \circ \mathcal{G}_{k_\ell} \circ \cdots \circ \mathcal{U} \circ \mathcal{G}_{k_1}  ,
	\end{equation}
	where $\mathcal{G}_{\mathrm{inv}} \coloneqq \mathcal{G}_{k_1}^\dagger\circ\cdots\circ\mathcal{G}_{k_\ell}^\dagger$ is the channel given by the composition of the inversion of all previous random gates channels.
\item Perform a POVM $E$ to be defined later and measure the \emph{survival probability}
	$F_{\mathbf{k}_\ell} = \Tr [E \ \mathcal C_{\mathbf{k}_\ell} (\rho)] $.
	To obtain an appropriate precision for $F_{\mathbf{k}_\ell} $, this step has to be repeated sufficiently often.
\item Repeat the previous step for sufficiently many (say $N$) random sequences  $\mathbf{k}_{\ell,1},\dots,\mathbf{k}_{\ell,N}$ of length $\ell$. Then, calculate the \emph{sequences survival probability}
	\begin{align}\label{eq:avg_seq_fidelity}
		F_{\rm{seq}}(\ell,\rho)=\frac{1}{N}\sum_{\mathbf{k}_\ell}F_{\mathbf{k}_\ell}=\frac{1}{N}\sum_{\mathbf{k}_\ell} \Tr [ E\,  \mathcal{C}_{\mathbf{k}_\ell}  (\rho) ] .
	\end{align}
	The number $K$ of random sequences should be chosen such that
	\begin{equation}
		F_{\mathrm{seq}} \approx F_{\mathrm{avg}} \ ,
	\end{equation}
	where $F_{\mathrm{avg}}$ is the survival probability averaged over all possible sequences. The choice can be motivated by an analysis on the variance of the random variable $F$, with $F_{\mathbf{k}_\ell}$ being a \emph{realization} and $F_{\mathrm{avg}}$ the \emph{mean} of the distribution. Note that, for a Clifford circuit, one requires a ``relatively small'' number of sequences to approximate $F_{\mathrm{avg}}$ \cite{RBFlammia,Helsen17}.
\item Repeat the previous steps for different lengths $\ell$.
\item Insert $F_{\mathrm{seq}}$ into the zeroth-order fitting model,
	\begin{equation}\label{eq:0_order_fitting_model}
	F_{\mathrm{avg}}^{(0)}(\ell, \rho)
	=
	\sum_{j=1} (\lambda_j\, d_j)^\ell \xi_j
	\end{equation}
	where the sum runs over the eigenvalues  $\set{d_j}_j$ of the target matrix $\mcU$ belonging to the space which $\rho$ has support over, with $\xi_j \coloneqq \Tr[ E \, \Lambda'(v_j)] \, \langle \rho , v_j \rangle$ absorbing the state preparation and measurement errors and where $\set{v_j}_j$ is the set of the basis vectors diagonalizing $\mcU$.
\item {Subsequently, we	 retrieve} the parameters $\{\lambda_j\}$ characterizing the average gate fidelity of $\widetilde \mcU$ with respect to  $\mcU$ (cfr. Appendix~\ref{sec:connection_avg_fidelity}) according to the relation
\begin{equation}\label{eq:relation_fidelity_and_traceI}
\EE({\mathcal F_{\Lambda^G, \mathcal I}})
 =
 \frac{\sum \lambda_j +d}{d(d+1)} ,
 \end{equation}
 {using classical estimation techniques.}
\end{enumerate}

 \paragraph*{The fitting model.}
 Considering a noise channel $\Lambda$ 
 {(we} now drop the subscript $\mcU$ to lighten notation) at each implementation of $\mcU \circ \mcG_k$, we can write
 \begin{equation}
 \mathcal{C}_{k_\ell} = \Lambda' \circ \mcG_{\mathrm{inv}}  \bigcirc_{t=\ell}^{1} \Lambda \circ \mcU  \circ \mcG_{k_t} .
 \end{equation}
 Note that the error channel $\Lambda'$ characterizing the implementation of $\mcG_{\mathrm{inv}}$ can be different from the error for the implementation of $ \mcU  \circ \mcG_{k_t} $.
 Now, defining recursively $\mcB_{k_t} \coloneqq \mcG_{k_t} \circ \mcB_{k_{t-1}}$ with $\mcB_{k_1} = \mcG_{k_1}$, and using the invariance of $U$ under the action of $G$, we can rewrite
 \begin{equation}
 \mathcal{C}_{k_\ell}
 =
 \Lambda' \bigcirc_{ t = \ell }^{1} \mcB_{k_t}^\dagger \circ \Lambda \circ \mcB_{k_t} \circ \mcU .
 \end{equation}
 When averaging over all possible sequences, we get
 $\mathcal{C}_{\mathrm{avg}}= \Lambda' \bigcirc_{t=\ell}^{1} \Lambda^G \circ \mcU $,
 where $\Lambda^G \coloneqq {|G|}^{-1} \sum_{j \in \mathbb N_{|G|}} \mcB_j^\dagger \circ \Lambda \circ \mcB_j $ is now like $U$ invariant with respect to the action of $G$.
 At this point we know that by Schur's Lemma (see Appendix~\ref{sec:Schur}) the Pauli-Liouville representations of $\mcU$ and $\Lambda^G$ are block-diagonalizable. If the decomposition of $G$ into irreducible representations does not contain any multiplicity for different the irreducible subspaces, the two matrices are simultaneously diagonalizable. If conversely multiple of the same irreducible representations occurs, in general there is no basis which brings both into a diagonal form and so, when diagonalizing $\mcU$, the matrix representation of $\Lambda^G$ will assume a block form, where each of these blocks corresponds to an irreducible representation.
 For the zeroth-order model, we consider the diagonal elements of $\Lambda^G$ only, and in particular the set $\set{\lambda_j}_j$ accounted in Eq.~\eqref{eq:0_order_fitting_model} denotes the ones belonging to the support of $\rho$. If one needs to take into consideration off-diagonal entries of $\Lambda^G$ too, that is, $O(\Delta t)$ terms, a first-order model is necessary {(see Appendix~\ref{sec:1_order_model})}.

\paragraph*{Statistical analysis and bootstrapping.} In order to estimate the average fidelity with the help of Eq.~\eqref{eq:relation_fidelity_and_traceI} we need to extract the decay rates $\set{\lambda_j}_j$ from the measured data. According to the protocol, this data consists of the survival probabilities $\{F_{\mathbf{k}_{\ell,q}}\}$ measured for $q=1,\dots, K$ randomly chosen sequences of lengths $\ell=1,\dots, \ell_{\max}$. Let us denote by $F_{\rm{seq}}(\ell)$ the average of $F_{\mathbf{k}_{\ell,q}}$ with respect to the randomly chosen symmetry-group sequences $k_{\ell,q}$ for fixed sequence length $\ell$. The zeroth-order fitting model from Eq.~\eqref{eq:0_order_fitting_model} then tells us that $F_{\rm{seq}}(\ell) \approx \sum_j (\lambda_j d_j)^\ell \xi_j$ and accordingly, we can
extract the parameters $\set{\lambda_j}_j$ from the sequence $\left(F_{\rm{seq}}(\ell)\right)_{\ell=1}^{\ell_{\max}}$ with the help of matrix pencil based signal reconstruction 
{(see Appendix~\ref{app:matrixPencilMeth})}. In order to improve our estimates we combine this approach with a bootstrapping procedure in the following way: To obtain a single bootstrap sample, we choose from the complete set of measured survival probabilities  $\{F_{\mathbf{k}_{\ell,q}}\}_{q=1,\ell=1}^{K,\ell_{\max}}$,  for each sequence length $\ell$ a random subset of the sequences $\mathbf{k}_{\ell,q}$ for which we then compute the average $F_{\rm{seq}}(\ell)$ and extract $\set{\lambda_j}_j$ as described before. Repeating this process for different random samples of the $\mathbf{k}_{\ell,q}$, we obtain different  estimates for the average fidelity, which is then averaged again with respect to the bootstrap samples in order to produce our final estimate for the average fidelity.

\begin{figure}[t!]
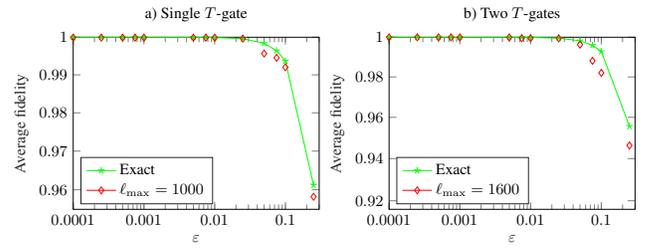

\centering
\includegraphics[width=0.23\textwidth]{SingleTgateL1000.tikz}
\includegraphics[width=0.23\textwidth]{twoTgateL1600.tikz}\\

 \caption{Average fidelity for noisy single {$T$}-gate (a) and two $T$-gates (b) for different noise-strengths $\varepsilon$ extracted from bootstrapping over  $N=100$ sequences of length up to $\ell=1000$ (a) and $\ell=1600$ (b). Green lines and stars indicate the analytic value of the average fidelity $\EE(\mc F)$ for the given noise level, whereas red diamonds the fitted average fidelity for the zeroth-order model.}
 \label{fig:TgateFit}
\end{figure}

\paragraph*{{Recovery} for single $T$-gate and two $T$-gates.}
In order to check the performance of our proposed protocol, we consider two examples: A single $T$-gate and the tensor product of two $T$-gates. In the single $T$-gate case, the generating Hamilonian is perturbed by an additional $\sigma_x$ term and we write $H = \frac{\pi}{8} \sigma_z - \varepsilon \sigma_x$, whereas in the two $T$-gates we consider  $ H = \frac{\pi}{8}\left(\sigma_z\otimes\1 + \1 \otimes\sigma_z - \varepsilon \, \sigma_x \otimes \sigma_x \right)$ for
varying  perturbation strength $\varepsilon$. Fig.~\ref{fig:TgateFit} compares the actual values of the average fidelity $\EE(\mc F)$, with the ones obtained by fitting the zeroth-order model for different values of $\varepsilon$ and different symmetry subspaces. The details of the fitting procedure are described in Appendix~\ref{sec:numerics:details}. From the two graphs it can be observed that we obtain very accurate predictions for the average fidelity already in the regime $\varepsilon \approx 0.25$.

\paragraph*{{Conclusions and outlook.}}
In this work, we have proposed a novel approach to randomized benchmarking, developing a protocol that allows to estimate individually the average gate fidelity of a unitary operator whose imperfect implementation is described by a gate-dependent noise channel, hence lifting some of the weaknesses related to the well-known protocols while {maintaining} robustness against SPAM errors. The key idea involves the twirl of the noise channel over the symmetry group of the target gate given by the composition of local symmetries with respect to the qubit subsystems and permutations thereof: This reduces considerably the amount of parameters that one has to retrieve and also the number of random gates needed. We have produced numerical simulations showing the validity of the protocol for a single-qubit and two-qubit $T$-gate, which proves to be faithful for relatively large noise levels. While we do not have an analytical formula for the scalability of the protocol, also due to the very wide range of possible gates that can be benchmarked in this way, the example of multiple tensor copies of the $T$-gate suggests that the number of non-linear parameters in the fitting model is drastically inferior compared to the full matrix dimension when increasing the number of qubits. Further investigations in this direction would be certainly of interest. Another open question concerns the number of different random samples needed to approximate the average fidelity with confidence, and in particular to obtain a bound in the fashion of Ref.~\cite{Helsen17}. With this letter, we hope to further stimulate research on randomized benchmarking outside the more established domain of the Clifford group and to explore the territory between randomized benchmarking and more traditional quantum process tomography.

 \paragraph*{Acknowledgements.} We thank R.\ Kueng, O.\ Buerschaper, R.\ Di Candia, {S.~T.~Flammia}, and M.\ Wageringel for valuable discussions.
This work has been supported by the ERC (TAQ), the DFG (EI 519/7-1, EI 519/11-1, CRC 183, {DAEDALUS}), {the excellence cluster MATH+,
and}
the Templeton Foundation. 
{This work has also received funding from the European Union's Horizon 2020
research and innovation programme under grant agreement No.~817482 (PASQuanS).}
AHW thanks the Humboldt Foundation 
for its support with a Feodor Lynen Fellowship. 

\setlength{\bibsep}{1pt}
\renewcommand{\bibfont}{\small}

\bibliographystyle{h-physrev}
\bibliographystyle{unsrt}

 \clearpage
\appendix

 	
 \section{Group theory}

 In this section, we provide the mathematical tools from group and representation theory which our benchmarking protocol relies upon; the reader already familiar with this mathematical field can skip the following introductory material.

 \begin{definition}[Group]
 	A \emph{group} $G$ is a set of elements equipped with a binary operation satisfying the following properties:\\
 	\emph{Closure}: For all $g,h \in G, \ g\cdot h \in G$.\\
 	\emph{Associativity}: For all $g,h,k\in G, \ (g\cdot h)\cdot k=g\cdot (h \cdot k)$.\\
 	\emph{Identity element}: There exist a unique identity element, $e$, such that for all $G \in G \ e \cdot g= g \cdot e = g$.\\
 	\emph{Inverse}: for every element $g \in G$ there exist an inverse element $g^{-1}$ such that $g^{-1} \cdot g= g \cdot g^{-1}= e$ .
 \end{definition}

 If two group can be linked by a  group isomorphism, they are said to be \emph{isomorphic}.
 They will then have many properties in common, in particular the same multiplication table or character table. One can therefore easily obtain information about a group if it is possible to find a group isomorphism connecting it to another well-known group; this is indeed what we do in our protocol to deal with the local symmetry groups.
 A subset $H \subset G$ is called \emph{subgroup} if all above conditions are still satisfied, e.g., the subset is closed with respect to the group operation. A subgroup $N$ such that $g^{-1} n \, g \in N$ for all $n \in N, \ g \in G$ is said to be \emph{normal} and this is denoted by $N \lhd G$.

 \medskip

 One can define a \emph{(left) group action of $G$ on a set $M$} by a function
 \begin{align}
 \phi: \quad G \times M &\rightarrow M ,\\
 (g,m) &\mapsto \phi (g,m)\nonumber
 \end{align}
 that fulfills the following two axioms:\\
 \emph{Identity}: for all $m \in M, \ \phi(e,m)=m$,\\
 \emph{Compatibility}: for all $m \in M, g,h \in G \ \phi(g,\phi(h,m))= \phi(g\cdot h, m)$.
 With this definition, we can furthermore define the following.
 \begin{definition}[Orbit]
 	An \emph{orbit} $G.m$ of an element $m \in M$ is given by all elements in $M$ obtained by the action of $G$, i.e.,
 	\begin{equation}
 	G.m \coloneqq \set{\phi (g,m) : g \in G} .
 	\end{equation}
 \end{definition}
 The action of $G$ on $M$ induces a \emph{partition} of the set $M$ itself, i.e.,
 it regroups the elements into subsets such that every element $m \in M$ is contained in one and only one of these.

 \begin{definition}[Stabilizer subgroup]
 	The \emph{stabilizer subgroup} of $G$ with respect to $m$ is the set of all elements on $G$ such that
 	\begin{equation}
 	G_m \coloneqq \set{ g \in G : \phi(g,m) =m } .
 	\end{equation}
 \end{definition}
 It is always possible to couple two groups to generate a new one. This is indeed what we looking for, having to combine symmetry of the local gates together with the invariance with respect to certain permutations thereof.
 \begin{definition}[Direct product]
 	Given two groups $G$ and $H$, the \emph{direct product} $G \times H$ is a \emph{Cartesian product} of ordered pairs $(g,h)$, with $g \in G, h \in G$ equipped with a binary operation acting component-wise, that is
 	\begin{equation}
 	(g_1,h_1) \cdot (g_2,h_2) = (g_1 \cdot g_2, h_1 \cdot h_2) .
 	\end{equation}
 \end{definition}
 This new structure satisfies all axioms of closure, associativity, existence of identity -- given by $(e_G,e_H)$ -- and inverse element-- $(g^{-1},h^{-1})$ being the inverse of $(g,h)$ -- and so it is a group.
 An alternative way to construct a new group from is given by the \emph{semi-direct product}.

 \begin{definition}[Outer semi-direct product]
 	Let $N$, $H$ be groups, $\varphi : H \rightarrow \mathrm{Aut} (N)$ be a homomorphism from $H$ to the set of automorphisms of $N$. Then the \emph{(outer) semi-direct product with respect to $\phi$}, denoted by $G=N \rtimes_\varphi H$, is the group whose underlying set are the pairs $(n,h) \in N \times H$ equipped with an operation defined as
 	\begin{align}
 	\bullet : G \times G  &\rightarrow G \\
 	((n_1,h_1),(n_2,h_2)) & \mapsto (n_1,h_1) \bullet (n_2,h_2) \\
 	&= (n_1 \cdot \varphi_{h_1}(n_2), h_1 \cdot h_2) , \nonumber
 	\end{align}
 	where $n_1,n_2 \in N, h_1,h_2 \in H$.
 \end{definition}
 This structure is again a group according to the defining axioms, with identity element $(e_N,e_H)$ and inverse $(n,h)^{-1}=(\varphi_{h^{-1}}(n^{-1}), h^{-1})$. Note that the the set $\{(n,e_H) : n \in N\}$ is a normal subgroup of $G$ isomorphic to $N$.

 \medskip

 It is also possible to go the other way around and obtain from a group $G$ and a normal subgroup $N$ a new group called \emph{quotient group}, denoted by $G/N$. This is the set of all cosets of $N$ in $G$, i.e.,
 \begin{equation}
 G/N \coloneqq \set{ g N : g\in G} ,
 \end{equation}
 where $g N$ is the \emph{left coset of $N$ in $G$} , namely
 \begin{equation}
 g N \coloneqq \set{ g n : n\in N} .
 \end{equation}
 The latter definitions stands for all subgroup $N$, not necessarily normal, however when $N$ is normal the left coset and the right coset (defined analogously) coincide.
 The set $G/N$ is then a group under the operation $(gN) \cdot (hN)=(gh)N$. We conclude this paragraph with the following definition of the \emph{canonical projection} which is involved in the construction of irreducible representations of a semi-direct product group, as we will see in the next subsection.
 \begin{definition}[Canonical projection]
 	Let $N \lhd G$. The group homomorphism
 	\begin{align}
 	\tau : G &\rightarrow G/N ,\\
 	g &\mapsto g N
 	\end{align}
 	is called \emph{canonical projection}.
 \end{definition}

 \section{Representation theory}

 We are now going to introduce representations, the core mathematical objects which the respective theory is named after.
 \begin{definition}[Representation]
 	A \emph{representation} of a group $G$ on a vector space $V$ is a group homomorphism onto the general linear group on $V$,i.e., a map
 	\begin{align}
 	\pi : H &\rightarrow \mathrm{GL}(V), \\
 	g &\mapsto \pi(g)
 	\end{align}
 	such that
 	\begin{equation}
 	\pi(g) \cdot \pi(h) = \pi (g \cdot h) .
 	\end{equation}
 \end{definition}
 A representation is said to be \emph{faithful} if it is injective, and its dimension corresponds to the dimension of the vector field $V$. A subspace $ W \subset V$ is said to be \emph{invariant} if, for all $g \in G$ and $w \in W$,
 \begin{equation}
 \pi(g)w \in W .
 \end{equation}
 Furthermore, a representation is said to be \emph{irreducible} if the only invariant subspaces are $\set{0}$ and $V$ itself; often, this is abbreviated as \emph{irrep}. Every complex representation of a finite group is \emph{completely reducible}, i.e., it can be decomposed as a direct sum of irreducible representations. This property, together with Schur's Lemma, makes irreducible representations and their \emph{characters} a central object in the theory and will also be particularly relevant in our work.
 \begin{definition}[Character of a representation]
 	The \emph{character $\chi_\pi$ of a representation $\pi$ of a group $G£$ on $V$} is given by
 	\begin{equation}
 	\chi_\pi (g) = \Tr [\pi (g) ] .
 	\end{equation}
 \end{definition}
 The dimension of a representation corresponds then to its character at the identity element, $\chi_\pi (e)$. For finite group, the number of irreducible representations is again finite, and the following result  is useful to check if all irreducible representations of a given group have been found.
 \begin{proposition}[Group order and irreducible representations dimension]
 	The order of a group $G$ and the dimension of its irreducible representations are linked by
 	\begin{equation}
 	|G|= \sum_{\alpha \ : \ \pi^{\alpha}  \mathrm{irrep}} \chi_{\pi^\alpha} (e) ^2 .
 	\end{equation}
 \end{proposition}
 One of the most important properties for character of irreducible representations is the following orthogonality relation.
 \begin{proposition}[Orthogonality formula]
 	Let $\set{\chi_\alpha}_\alpha$ be the set of characters of all irreducible representations of a group $G$.
 	Then
 	\begin{equation}
 	\frac{1}{|G|} \sum_{g \in G} (\chi_\alpha(g))^\ast  \chi_\beta (g)
 	=
 	\begin{cases}
 	1 & \text{if } \alpha = \beta \\
 	0 & \text{if } \alpha \neq \beta
 	\end{cases}
 	\end{equation}
 \end{proposition}
 From this, follows one of the key results in representation theory is the formula for multiplicities, used to decompose a representation into its irreducible components.
 \begin{proposition}[Multiplicity formula]
 	Let $\chi_\alpha$ be the character of the irreducible representation $\pi^\alpha$ and $\phi$ the character of the representation $\pi$ of a group $G$.
 	Then
 	\begin{equation}
 	\frac{1}{|G|} \sum_{g \in G} (\chi_\alpha (g))^\ast \phi (g)
 	=
 	m_\alpha ,
 	\end{equation}
 	where $m_j$ is the multiplicity of the irreducible representation $\pi_j$ in the decomposition of $\pi$, so that $\pi$ is similar
	to a block diagonal matrix in the form
 	\begin{equation}
 	\pi (g) \simeq  \bigoplus \pi_\alpha (g) \otimes \1_{m_\alpha} \qquad \forall g \in G ,
 	\end{equation}
 	with $\1_{m_j}$ being the identity matrix on $\mathds{C}^{m_j}$.
 \end{proposition}

 For each irreducible subspace of $V$ we can choose a basis set $\set{v_j^\alpha}_j$ where $\alpha$ is a label of the irreducible representation (with dimension $\mathrm{dim} \, \alpha$) decomposing $\pi$ and $j \in \set{1,\dots ,\mathrm{dim} \, \alpha}$. Each vector $v$ in $V$ can then be written as their linear combination $v = \sum_{\alpha} \sum_{j=1}^{\mathrm{dim} \, \alpha} c_j^\alpha \, v_j^\alpha$. We can conversely identify the basis vector components of any vector $v$ by application of an appropriate projector $\mathrm{P}_j^\alpha$, such that $\mathrm{P}_j^\alpha\, v = c_j^\alpha \, v_j^\alpha$, where
 \begin{equation}\label{eq:projectors_onto_irred_basis}
 	\mathrm{P}_j^\alpha
 	=
 	\frac{\mathrm{dim} \, \alpha}{\left|G\right|} \sum_{g \in G} (\pi^\alpha (g))_{{j,j}}^\ast \, \pi(g).
 \end{equation}
 Note that, in order to construct these projectors, the knowledge of the sole diagonal elements of the corresponding irreducible representation $\pi^\alpha$ is sufficient. By having access to the character table only, it is still possible to project any vector onto distinct irreducible subspaces (up to multiplicity) by using
 \begin{equation}\label{eq:projectors_onto_irred_subspace}
 \mathrm{P}^\alpha
 =
 \frac{\mathrm{dim} \, \alpha}{\left|G\right|} \sum_{g \in G}   (\chi_\alpha (g))^\ast \, \pi(g).
 \end{equation}

 \section{Pauli-Liouville representation}

 To represent density operators and quantum channel on $n$ qubits as vectors and matrices respectively, we will make use of the \emph{Pauli-Liouville representation} with respect to Pauli basis. Let us pick
 \begin{equation}
 B \coloneqq \left\{\frac{1}{\sqrt 2 ^n} \bigotimes_{\nu=1}^n \tilde \sigma_\nu : \tilde \sigma_\nu \in \set{\1_2,X,Y,Z}\right\} ,
 \end{equation}
 where $X,Y,Z$ are the Pauli matrices, as an orthonormal basis of $\mathrm{GL} (\mathds{C}^{2n})$ with respect to the Hilbert-Schmidt inner product $\langle A, B \rangle \coloneqq \Tr [A^\dagger B]$. Then we know that we can express any density operator $\rho$ and quantum channel $\mathcal{C}$ respectively as
 \begin{equation}
 \rho= \sum_{j=1}^{4^n} \rho_j \sigma_j \qquad \text{and} \qquad \mathcal C (\rho) = \sum_{j=1}^{4^n} \mathcal C (\sigma_j) \rho_j ,
 \end{equation}
 where $\sigma_j \in B$ and $\rho_j\coloneqq \langle \sigma_j, \rho \rangle$, so that we can represent them as
 \begin{equation}
 \ket{\rho} = \begin{pmatrix} \rho_1 \\ \rho_2 \\ \dots \\ \rho_{4^n} \end{pmatrix} \qquad \text{and} \qquad \mcC_{j k} = \langle \sigma_j , \mathcal{C} (\sigma_k) \rangle .
 \end{equation}
 In this way, we may represent $\mathcal{C}(\rho)$ as a matrix-vector multiplication $\mcC \ket{\rho}$ and the concatenation of two channels $\mathcal{D}$ and $\mathcal{C}$ as a matrix multiplication $\mcD \,  \mcC$. Additionally, representing channels in matrix form will allow us to make use of Schur's Lemma for matrix representations (see section~\ref{sec:Schur}).
 We can analogously represent a POVM $E$ in the form
 \begin{equation}
 \bra{E} = \left( E_1 \ E_2 \dots E_{4^n} \right) \qquad \text{with} \qquad E_j=\langle E , \sigma_j \rangle .
 \end{equation}
 With this, the probability to obtain an outcome described by $E$ when measuring $\rho$ is $p(E|\rho)= \Tr [E \rho ] = \langle E, \rho \rangle$.

 \section{Zassenhaus Formula}\label{sec:Zassenhaus}

 In order to justify our mathematical assumption through physical motivations, let us consider that the gate $\widetilde{U}$, which is the physical realization of the ideal gate $U$, is obtained during the application of some Hamiltonian $H$, perturbed for a fraction of time $\Delta t$ (we denote the perturbed Hamiltonian as $R$), i.e.,
 $\widetilde{U}=\e^{-\i (R \Delta t +H T)}$. Using the \emph{Zassenhaus formula} \cite{Zassenhaus}, we can rewrite the implemented gate as
 \begin{align}
 \widetilde{U}
 &=
 \e^{-\i H T}  \e^{-\i R \Delta t} \prod_{n=2}^\infty e^{C_n(H T, R \Delta t)} \\
 &=
 \e^{-\i H T}  \left( \1 -\i R \Delta t +  \S \Delta t \right) +O(\Delta t^2) ,
 \end{align}
 where
 \begin{equation}
 \S \coloneqq  \sum_{n=2}^\infty c_n
 [ \underbrace{ H,[H,\dots,[H }_{n-1 \text{ times}}  ,R] \dots]]  T^{n-1} ,
 \end{equation}
 with the Zassenhaus coefficients $c_n$ that can be recursively calculated as for instance in Ref.~\cite{Zassenhaus}.
 This implies that the off-diagonal elements of the matrix $\widetilde{U}$ -- computed in the eigenbasis of $U$ -- are of order $\Delta t$, justifying our assumption on the noise $\Lambda$.

 \section{First-order fitting model}\label{sec:1_order_model}
 We will now take into account off-diagonal matrix entries. The main reason of the protocol is that, by twirling the error channel over the symmetry group $G$, we reduce the number of these off-diagonal matrix entries and so we drastically decrease the amount of parameters in the fitting model. Let us write $\Lambda^G = \Lambda_0+\Lambda_{\mathrm{off}}$, with $\Lambda_0$ being jointly diagonalizable with $\mcU$. Provided $\Lambda_{\mathrm{off}}=\set{\mu_{i,j}}_{i\neq j}$ to be ``small'' (i.e., the second order perturbation being negligible), we can consider the first-order model
 \begin{equation}
 F_{\mathrm{avg}}^{(1)}(\ell, \rho)
 =
 F_{\mathrm{avg}}^{(0)} (\ell, \rho) +
 \sum_{i \neq j} \sum_{p=0}^{\ell-1} (\lambda_i d_i)^p\, (\lambda_j d_j)^{\ell-p-1} \, \zeta_{i,j} ,
 \end{equation}
 with $\zeta_{i,j} \coloneqq \mu_{i,j} d_j\ \,\Tr[E \, \Lambda'(v_i)] \langle \rho , v_j \rangle $ and the indices $i,j$ labeling the elements within the support of $\rho$. This expression may be re-formulated into a simpler form, e.g., using the geometric series formula we obtain
 \begin{equation}
 F_{\mathrm{avg}}^{(1)} (\ell, \rho)
 =
 F_{\mathrm{avg}}^{(0)} (\ell, \rho)
 +
 \sum_{i\neq j} \frac{(\lambda_j d_j)^\ell- (\lambda_i d_i)^\ell}{\lambda_j d_j- \lambda_i d_i} \zeta_{i,j} .
 \end{equation}
 As already mentioned, since we twirled over the symmetry group and so $\Lambda^G$ is block-diagonal, a number of $\mu_{i,j}$ (and hence the corresponding $\zeta_{i,j}$) can be set to $0$ in advance. More precisely, when a representation of the symmetry group is written as a direct sum of irreducible representations as
 \begin{equation}
 \pi (g)=\bigoplus_{\alpha \, \mathrm{irrep}} \1_{m_\alpha} \otimes \pi^{\alpha}(g) ,
 \end{equation}
 where $m_\alpha$ is the multiplicity of the irreducible representation $\pi^\alpha$, two matrices $X$ and $Y$
 which are both commuting with $\pi (g)$ assume the form
 \begin{align}\label{special_matrix_form}
 X= \bigoplus_{\alpha} x^\alpha \otimes \1_{\mathrm{dim} \, \alpha}
 && \text{and} &&
 Y= \bigoplus_{\alpha} y^\alpha \otimes \1_{\mathrm{dim} \, \alpha} ,
 \end{align}
 where $x^\alpha, y^\alpha$ are square matrices with $\mathrm{dim} \, x^\alpha = \mathrm{dim}\,  y^\alpha = m_\alpha$.
 If $X$ is normal, one can then choose a basis such that all $x^\alpha$ are diagonal (so that $X$ will assume a diagonal form), while $Y$ will maintain a similar form $	Y= \bigoplus_{\alpha} \widetilde{y}^\alpha \otimes \1_{\mathrm{dim} \, \alpha}$. Hence, in our case, while diagonalizing $\mcU$ (from the Pauli-Liouville representation), $\Lambda^G$ maintains a form as in Eq.~\eqref{special_matrix_form}.

 \section{Construction of irreducible representations of semi-direct product groups}\label{sec:direct_product_irreducible representations}

 As we have discussed in the previous section, it is possible to couple two groups to construct a new one using direct and semi-direct products.  We can also obtain all irreducible representations of the latter using knowledge about irreducible representations of the original two groups alone. For direct {products}, the procedure is {straightforward}.

 \begin{theorem}[\cite{Serre}, Theorem 10, Chapter 3]
 	Each irreducible representation of a direct group $G_1 \times G_2$ is isomorphic to a representation $\pi_1 \otimes \pi_2$ with $\pi_1$ and $\pi_2$ being irreducible representations of group $G_1$ and $G_2$ respectively
 \end{theorem}

 For a group generated by a semi-direct product $N \rtimes H$, a more sophisticated machinery is needed (cfr. Refs.~\cite{Serre,Berndt}), and works only if the normal subgroup $N$ is also \emph{abelian}, i.e., all elements commute with respect to the group operation.
 Assuming $N$ to be abelian, its irreducible representations $\set{\chi_\alpha}_\alpha$ are 1-dimensional and carry an action of $G$ by
 \begin{equation}
 g  \cdot \chi_\alpha (a) = \chi_\alpha (g^{-1} a g) \qquad \forall a \in N \text{ and } g \in G .
 \end{equation}
 Now, consider the orbits of the characters induced by the action of $H$ and choose a set of representatives $\set{\chi_r}_r$. For each $r$, let $H_r$ be the stabilizer subgroup of $\chi_r$ in $H$ and then define $G_r=G_0 \cdot H_r$. Now extend $\chi_r$ to $G_r$ by
 \begin{equation}\label{eq:extend_character}
 \chi_r (ah)= \chi (a) \qquad \forall a \in N \text{ and } h \in H_r .
 \end{equation}
 Let $\theta$ be an irreducible representations of $H_r$ and lift it to an irreducible representation $\widetilde{\theta}$ of $G_r$ through the canonical projection $P: G_r \rightarrow G_r / N$. As a final step, compose the two representations and obtain a representation $\rho_{r,\widetilde{\theta}}$ of the group $G$ by \emph{induction}, i.e., $\rho_{r,\widetilde{\theta}}= \mathrm{Ind}_{G_r}^G(\chi_r\cdot \widetilde{\theta})$. From Ref.\ \cite[Proposition 25]{Serre}, we know that the so constructed representations $\rho_{r,\widetilde{\theta}}$ are irreducible and exhaust all irreducible representations of $G$.
 Since we will only need the characters $\chi_{\rho_{r,\widetilde{\theta}}}$ of
 the irreducible representations of $G$ to apply Schur's Lemma and so apply our protocol, we will not elaborate on what induced representations are. To obtaind the sought characters, it suffices to make us of a Mackey-type formula
 \begin{equation}\label{eq:induced_character}
 \chi_{\rho_{r,\widetilde{\theta}}} (s)
 =
 \frac{1}{|G_r|} \sum_{\substack{g\in G \\ g^{-1} s g \in G_r}}
 \chi_r \cdot \chi_{\widetilde{\theta}}\  (g^{-1} s g) .
 \end{equation}

 \section{Schur's Lemma}\label{sec:Schur}

 Hereby we write one of the most important results in representation theory, namely \emph{Schur's Lemma}. We will restrict it on finite-dimensional  representations case.
 \begin{lemma}[Schur's Lemma]
 	Let $\pi_\alpha$ and $\pi_\beta$ be two irreducible representations of a finite group $G$ of dimension $m$ and $n$ respectively, and $M$
	an $m \times n$ matrix.
 	If
 	\begin{equation}
 	\pi_\alpha (g)\, M\, \pi_\beta^{-1}(g)=M  \qquad \forall g\in G
 	\end{equation}
 	then $\pi_\alpha$ and $\pi_\beta$ are equivalent irreducible representations or $M=0$.
 	
 	\smallskip
 	
 	\noindent Furthermore, if
 	\begin{equation}
 	\pi_\alpha(g)\, M\, \pi_\alpha^{-1} (g)=M  \qquad \forall g\in G
 	\end{equation}
 	then $M=\mu \mathds 1$, i.e., it is a scalar matrix.
 \end{lemma}

 \section{Connecting to the average gate fidelity}\label{sec:connection_avg_fidelity}

 For a quantum channel $\mathcal E$ and a unitary operation $\mcU$, the \emph{gate fidelity} between these two quantities for a pure state $\phi$ is given by
 \begin{equation}
 	\mathcal F _{\mathcal E, \mcU} (\phi)
 	\coloneqq
 	\Tr\left[ \mcU (\ket{\phi}\bra{\phi}) \, \mathcal E (\ket{\phi}\bra{\phi}) \right]
 \end{equation}
 and defining $\Lambda = \mcU^\dagger \circ \mathcal E$ one has
 \begin{equation}
 	\mathcal F _{\mathcal E, \mcU} (\phi)=\mathcal F _{\Lambda, \mathcal I} (\phi)
 	=
 	\Tr\left[ \ket{\phi}\bra{\phi} \, \Lambda (\ket{\phi}\bra{\phi}) \right] ,
 \end{equation}
 hence quantifying the noise channel $\Lambda$ for the implementation $\mathcal E$ of $\mcU$.\\
 The \emph{average gate fidelity} is then obtained by integrating this quantity over the Haar measure on pure states, that is,
 \begin{equation}
\E({\mathcal F _{\mathcal E, \mcU}})= \EE({\mathcal F_{\Lambda, \mathcal I}})
 \coloneqq
 \int_{\mathrm{Haar}} \Tr\left[ \ket{\phi}\bra{\phi} \, \Lambda (\ket{\phi}\bra{\phi}) \right] \, \mathrm d \phi.
 \end{equation}
 Conversely, the \emph{entanglement fidelity} of a quantum channel $\mathcal{E}$, defined as
 \begin{equation}
 F_{\mathrm{ent}}(\mathcal{E})
 \coloneqq
 \bra{\psi} (\mathcal{I}\otimes \ \mathcal{E})(\ket{\psi}\bra{\psi})\ket{\psi} ,
 \end{equation}
 with $\ket{\psi}$ being a maximally entangled state vector, can be written as~\cite{Nielsen02}
 \begin{equation}
 	F_{\mathrm{ent}} (\mathcal E) = d^{-3} \sum_j \Tr[ V_j^\dagger \mathcal{E} (V_j) ] ,
 \end{equation}
 for any orthonormal basis $\set{V_j}_j$ such that $\Tr[V_j V_k] = d \, \delta_{j,k}$ (in the case of $n$ qubits, $d=2^n$).
The average gate fidelity of $\mathcal E$ is then linked to this quantity by~\cite{Nielsen02}
 \begin{equation}\label{eq:fidelity_formula}
\EE({\mathcal{F}_{\mathcal E, \mathcal I}})
 =
 \frac{d F_{\mathrm{ent}}(\mathcal{E}) +1}{d+1}
 =
 \frac{\sum_j \Tr[ V_j^\dagger \mathcal{E} (V_j) ] +d^2}{d^2(d+1)} ,
  \end{equation}
  so that the average gate fidelity of the twirled error channel $\Lambda^G$ is related to the parameters $\set{\lambda_j}_j$ obtained in the fitting model in Eq.~\eqref{eq:0_order_fitting_model} by
 \begin{equation}\label{eq:relation_fidelity_and_trace}
\EE({\mathcal F_{\Lambda^G, \mathcal I}})
 =
 \frac{\sum \lambda_j +d}{d(d+1)} .
 \end{equation}
 Now the question is what information about the original noise channel we can extract from the twirled channeld $\Lambda^G$. In fact they are the same, since the entanglement fidelity is invariant under twirling over the symmetry group $G$. Let us rewrite
 \begin{align}
 F_{\mathrm{ent}}(\Lambda^G)
 &=
 d^{-3} \sum_j \Tr[ V_j^\dagger \Lambda^G (V_j) ] \\
 &=
 \frac{d^{-3}}{|G|} \sum_{k=1}^{|G|} \sum_j \Tr[ V_j^\dagger g_k^\dagger \Lambda (g_k V_j g_k^\dagger)g_k ] \\
 &=
 \frac{d^{-3}}{|G|} \sum_{k=1}^{|G|} \sum_j \Tr[ (W_j^k)^\dagger \Lambda (W_j^k) ] ,
 \end{align}
 where we denote $W_j^k= g_k V_j g_k^\dagger$ and used cyclicity of the trace. Since $W_j^k$ is again an orthogonal basis with respect to the Hilbert-Schmidt inner product (i.e., $\Tr [(W_{j'}^k)^\dagger W_j^k]= d \,\delta_{j'j} \  \forall k$),  then $d^{-3} \sum_j \Tr[ (W_j^k)^\dagger \Lambda (W_j^k) ]= F_{\mathrm{ent}}(\Lambda)$ so that
 \begin{equation}
 F_{\mathrm{ent}}(\Lambda^G) = \frac{1}{|G|} \sum_{k=1}^{|G|} F_{\mathrm{ent}} (\Lambda) = F_{\mathrm{ent}} (\Lambda)
 \end{equation}
 and hence
 \begin{equation}
\EE({\mathcal F_{\Lambda, \mathcal I}})=
\EE({\mathcal F_{\Lambda^G, \mathcal I}} ).
 \end{equation}

 \section{Characterizing the error of the single gate $U$}

 In order to recover the fidelity of the gate $U$ from the noise $\Lambda$, which originates from the composition of $U$ and a unitary gate from the symmetry group $G$, we first consider the \emph{$\chi$ matrix representation of $\mathcal{E}$},
 \begin{equation}
 \mathcal{E}(\rho) = \sum_{i,j} \chi_{i,j} P_i \rho P_j \ .
 \end{equation}
 We can characterize the error of the gate $U$, distinguishing it from the one coming from the symmetry group $G$, that we consider to be ${\mathcal{N}}$ for all element in the group (which can be benchmarked separately using for instance the known methods to benchmark Clifford gates), using the bound from Ref.~\cite[Appendix D]{KimmOhki} (where we set $i=0$)
 \begin{align}
 |\chi_{0,0}^{\Lambda \circ \mathcal{N}} -\chi_{0,0}^\Lambda\chi_{0,0}^{\mathcal{N}}|
 &\leq
 2\left({(1-\chi_{0,0}^\Lambda)\chi_{0,0}^\Lambda(1-\chi_{0,0}^{\mathcal{N}})\chi_{0,0}^{\mathcal{N}}} \right)^{1/2}
 \nonumber \\
 &+
 (1-\chi_{0,0}^\Lambda)(1-\chi_{0,0}^{\mathcal{N}}).
 \end{align}
 For an arbitrary channel $\mathcal{E}$, we know that $\chi_{0,0}^{\mathcal{E}} = \Tr[\mathcal{E}] / d^2$ (cfr.\cite[Eq.\ (2.30)]{MagGamEmer} and Eq.~\eqref{eq:relation_fidelity_and_trace}), so that we can recover the fidelity of the gate $U$ form the ones of the gates belonging to $G$ and from $\EE({\mathcal F_{(\Lambda\circ \mc N)^G, \mathcal I}})$ obtained with our protocol.
 The bound in particular is valid in the regime $\chi_{0,0}^{\mathcal{N}} \approx 1$, i.e., when the  gates of the symmetry group can be implemented with high fidelity.

\section{Confidence interval}
{This section is concerned with uncertainty quantification in our scheme.}
To {assess} the number of different random sequences that have to be sampled in order to justify $F_{\rm seq}(\ell) \approx F_{\rm avg}(\ell)$ for a given sequence length $\ell$, Wallmann and Flammia in Ref.~\cite{RBFlammia} provided bounds on the variance for the Clifford randomized benchmarking protocol described in Ref.~\cite{MagGamEmer}. Their results show that a relatively small number of random sample is needed. We want to show a bound similar to that of Ref.\ \cite[Theorem 10]{RBFlammia},
\begin{equation}
	\sigma_\ell^2
	=
	\frac{1}{|G|^\ell} \sum_{\mathbf{k}_\ell} F_{\mathbf{k}}^2(\ell,\rho)-	F_{\operatorname{avg}}(\ell,\rho)^2 .
\end{equation}
In Pauli-Liouville representation and using $(E|\mathcal{C}|\rho)^2=(E^{\otimes 2}| \mathcal{C}^{\otimes 2} | \rho^{\otimes 2})$, this can be expressed in terms of a scalar product in the form
\begin{equation}
	\sigma_\ell^2
	=
	\frac{1}{|G|^\ell} \sum_{\mathbf{k}_\ell}(E^{\otimes 2}| \mathcal{C}_{\mathbf{k}_\ell}^{\otimes 2} | \rho^{\otimes 2})-	
	(E^{\otimes 2}| \mathcal{C}_{{\rm avg},\ell}^{\otimes 2} | \rho^{\otimes 2}) .
\end{equation}
Now, we assume to be in the regime $\Lambda=\1+Q\Delta t$, where $Q$ is a bounded matrix under additional assumption $\Tr Q=\Theta (d^2)$, and expand the expression for the variance up the second order in $\Delta t$.
\begin{widetext}
\begin{equation}\label{eq:variance}
	\sigma_\ell^2
	=
	\Delta t^2(E^{\otimes 2}|
	\sum_{j=1}^\ell \frac{1}{|G|} \sum_{B \in G}  (\mcU \otimes \mcU)^{\ell-j} (\mcB ^\dagger Q \mcB  \otimes \mcB^\dagger Q \mcB ) (\mcU \otimes \mcU)^{j} -  \sum_{j=1}^\ell (\mcU \otimes \mcU)^{\ell-j} (Q^G \otimes Q^G) (\mcU \otimes \mcU)^{j}
	|\rho^{\otimes 2} )\\
	+O(\ell^2 r^2 d^4) .
\end{equation}
\end{widetext}
The first term can be bounded as in Ref.\ \cite{RBFlammia} using diamond norm properties and Ref.\ \cite[Proposition 9]{RBFlammia} with $4d(d+1)\ell r$. Again following that argument, the terms of order $O(\Delta t^3 Q^3)$ and $O(\Delta t^4 Q^4)$ are $O(\ell^2 r^2 d^4)$.
Knowing the structure of $Q^G$ from the analysis of the symmetry group $G$, we can upper bound the number of non-zero terms as
\begin{equation}
	\sum_{\alpha} m_\alpha^2 \, d_\alpha \leq \max_{\alpha} m_\alpha \sum_{\alpha} m_\alpha \, d_\alpha = \max_\alpha m_\alpha \, d^2 .
\end{equation}
From now on, we denote $m=\max_\alpha m_\alpha$ and $q=\max_{i,j} q_{i,j}$, the largest matrix entry of $Q$  {that} we assume being {independent} of $d$.
The second term in expression~\eqref{eq:variance} obeys to the inequality
\begin{equation}\label{eq:variance_2_term}
	(E^{\otimes 2}|
	\sum_{j=1}^\ell (\mcU \otimes \mcU)^{\ell-j} (Q^G \otimes Q^G) (\mcU\otimes \mcU)^{j}
	|\rho^{\otimes 2} )
	\leq
	\ell q^2  m^2  d^4 .
\end{equation}
Using
\begin{equation}
	\Tr[\Lambda]=d(d+1)\EE (\mc F) - d
\end{equation}
follows
\begin{equation}
	\Delta t=-r\frac{d(d+1)}{\Tr[Q]} ,
\end{equation}
and so $\Delta t=O(r)$ since we assumed $\Tr Q=\Theta (d^2)$. Hence, second term of Eq.~\eqref{eq:variance} is $ O(m^2\,\ell\, r^2 d^4)$. While we do not have an exact estimation for the scaling of $m$ for the general case, in the illustrated example for tensor copies of $T$ gate this goes as $O(\log d)$.
Summarizing gives a bound for the variance as
\begin{equation}
	\sigma^2_\ell \leq 4d(d+1)\ell r+O(\ell^2 r^2 d^4)+ O(m^2\,\ell\, r^2 d^4) ,
\end{equation}
 where the second term dominates the third one for $\ell \gg m^2$, i.e., in this regime the bound is exactly equivalent to the one of
 Ref.\
 \cite[Theorem 10]{RBFlammia}.
 This bound however is probably not tight, and we are interested whether a bound similar to the one provided in Ref.~\cite{Helsen17} can be obtained.

 \pagebreak

 \section{Example {of} tensor copies of $T$-{gates}}
 
 {In this section we present an example that 
shows the functioning of our scheme.}
 We present an example to assess our protocol on one of the most relevant quantum gates, the $T$ gate, which together with the $H$, $S$ and CNOT gates gives rise to a universal quantum circuit. {Specifically within the context of fault tolerant quantum computing, this situation is of
 paramount importance.} 
 We are going to benchmark tensor copies of this gate too, up to four, in order to get a feeling on the scalability and necessary resources for this method.
 We give in the following the step-by-step sequence.
 \begin{enumerate}[label={[\arabic*]}]
 	\item Produce the $n$-Kronecker product group, denoted by $A_n$, of the local abelian symmetry group
 	\begin{equation*}
 		\left\{
 		\left(
 		\begin{array}{cccc}
 		1 & 0 & 0 & 0 \\
 		0 & 1 & 0 & 0 \\
 		0 & 0 & 1 & 0 \\
 		0 & 0 & 0 & 1 \\
 		\end{array}
 		\right),
 		\left(
 		\begin{array}{cccc}
 		1 & 0 & 0 & 0 \\
 		0 & 0 & -1 & 0 \\
 		0 & 1 & 0 & 0 \\
 		0 & 0 & 0 & 1 \\
 		\end{array}
 		\right), \right.
 		\end{equation*}
 		\begin{equation*} \left.
 		\left(
 		\begin{array}{cccc}
 		1 & 0 & 0 & 0 \\
 		0 & 0 & 1 & 0 \\
 		0 & -1 & 0 & 0 \\
 		0 & 0 & 0 & 1 \\
 		\end{array}
 		\right),
 		\left(
 		\begin{array}{cccc}
 		1 & 0 & 0 & 0 \\
 		0 & -1 & 0 & 0 \\
 		0 & 0 & -1 & 0 \\
 		0 & 0 & 0 & 1 \\
 		\end{array}
 		\right)
 		\right\},
 	\end{equation*}
 	which is isomorphic to the \emph{cyclic group} of order 4, $\Z_4$.
 	\item Construct representation of the symmetric group $S_n$ permuting the local subsystems.
 	\item Construct the full symmetry group $G$ as a semi-direct product of $A_n$ and $S_n$ by multiplying the respective $4^n$-dimensional matrix representations. Each $g \in G$ is given by
 	$g=a.\sigma$, with $a\in A_n, \sigma \in S_n$.
 	\item From the character table of $\Z_4$,
 	\begin{center}
 		\begin{tabular}{l|c|c|c|c|}
 			$\Z_4$& $e$ & $\gamma$ & $\gamma^2$ & $\gamma^3$ \\
 			\hline
 			$\chi_0$ & 1 &1 &1 &1 \\
 			$\chi_1$ & 1 &$\i$ &-1 &-$\i$ \\
 			$\chi_2$ & 1 &-1 &1 & -1 \\
 			$\chi_3$ & 1 &-$\i$ &-1 &$\i$
 		\end{tabular}
 	\end{center}
 	construct the character table of $A_n$ by taking the product of the respective characters \begin{equation}
 	\chi_{c_1,c_2,\dots,c_n}(\ell_1,\ell_2,\dots,\ell_n) \coloneqq \chi_{c_1}(\ell_1) \chi_{c_2}(\ell_2) \dots  \chi_{c_n}(\ell_n)\ ,
 	\end{equation}
 	where $\ell_j \in \Z_4$ and $c_j$
 	is the label representing the irreducible representation.
 	\item Divide the characters of $A_n$ into orbits with respect to the action of $S_n$ given by
 	\begin{equation}
 	\sigma.\chi(a)_{c_1,c_2,\dots,c_n} \coloneqq \chi(\sigma^{-1} a\,  \sigma)_{c_1,c_2,\dots,c_n} .
 	\end{equation}
 	In this particular case, the action of $S_n$ works as a permutation of the labels of the irreducible representations, i.e.,
 	\begin{equation}
 	\sigma.\chi(a)_{c_1,c_2,\dots,c_n} =\chi(a)_{\sigma(c_1,c_2,\dots,c_n)}.
 	\end{equation}
 	Choose for each orbit a representative element, for instance $\chi(a)_{c_1,c_2,\dots,c_n}$ with $c_1\leq c_2 \leq \dots \leq c_n$, building a set $\set{\chi_j}_j$.
 	\item For each representative element $\chi_j$, find the stabilizer group $H_j$ as a subgroup of $S_n$.
 	\item For each irreducible representation $\pi$ of $H_j$, write an irreducible representation of the subgroup $G_j\coloneqq A_n \cdot H_j$ of $G$ by
 	\begin{equation}
 	\tilde{\rho}^j_\pi (a, g_j)  = \chi_j (a)\cdot \pi (g_j) .
 	\end{equation}
 	\item Obtain the characters of the representation $\rho^j_\pi$ of $G$ induced by $\tilde{\rho}^j_\pi$ with the Mackey-type formula,
 	\begin{equation}\label{eq:induced_character_2}
 	\chi_{\rho^j_\pi} (s)
 	=
 	\frac{1}{|G_j|} \sum_{\substack{t\in G \\ t^{-1} s t \in G_j}}
 	\chi_{\tilde{\rho}^j_\pi}  (t^{-1} s t) ,
 	\end{equation}
 	and obtain the irreducible representation multiplicity $m^j_\pi$ in the decomposition of the Pauli-Liouville representation of $G$ by the formula
 	\begin{equation}
 	m^j_\pi = \frac{1}{|G|} \sum_{g \in G} 	\left(\chi_{\rho^{j}_\pi} (g) \right)^{\ast} \cdot \phi (g) ,
 	\end{equation}
 	where $g\mapsto \phi(g)$ is the trace of $g$ in Pauli-Liouville representation.
 \end{enumerate}

 In case of $n=4$, for instance, there are 256 different irreducible representations of $A_4$ and five stabilizer groups: the full permutation group $S_4$ for the irreducible representations of the form $\chi_{a,a,a,a}, \ a \in {0,1,2,3}$, giving rise to $4\cdot 5=20$ irreducible representations for $G$, $S_3$ for the representative irreducible representations of the form $\chi_{a,a,a,b}$ and $\chi_{a,b,b,b}$ with $a<b$, giving rise to $12\cdot 3=36$ new irreducible representations, $S2\times S2$ (isomorphic to the Klein 4 group) for representative elements $\chi_{a,a,b,b}$ with $a<b$, so that a total of $6\cdot 4=24$ irreducible representations of $G$ are derived,  again representative elements $\chi_{a,a,b,c},\chi_{a,b,b,c},\chi_{a,b,c,c}$ with $a<b<c$ have stabilizer $S_2$, producing additional $12\cdot 2=24$ irreducible representations; finally, the single representative element $\chi_{0,1,2,3}$ is the representative element of the sole orbit with trivial stabilizer. Hence, we have in total 105 different induced irreducible representations of $G$ whose characters is obtained using Eq.~\eqref{eq:induced_character_2}. As one can see from Table
 {IV}
 only 22 of these irreducible representations decompose the twirled noise matrix, and the trivial representation has the highest multiplicity.

 \section{Results for $n \leq 4$}

 \begin{table*}
		{\begin{tabular}{|c|c|}
 				\hline
 				Irreducible representation & Multip.  \\
 				\hline
 				$\chi_0$ & 2\\
 				$\chi_1$ &1 \\
 				$\chi_3$ & 1\\
 				\hline
 		\end{tabular}
		\caption{This and the subsequent three tables depict the irreducible decompositions 
		of the symmetry group $G$  of multiple tensor copies of the $T$-gate channel. 
		The superscripts of $\chi$ label the irreducible representations of $A_n$, the word after the semicolon
 		the irreducible representation of the stabilizer group: $e$ denotes the trivial representation, $\rm{sgn}$ the sign representation, $\rm{std}$ the standard representation, $\rm{ker}_a$ the \emph{Kernel $a$} representation of the Klein 4 group, $2\rm{dim}$ the 2-dimensional representation of $S_4$.} \label{tab:irreducible representation_decomposition_T_gate. This first table shows the one $T$-gate decomposition.}}
\end{table*}
 \begin{table*}

		\begin{tabular}{|c|c|c||c|c|c|}
 				\hline
 				Irreducible representation & Dim & Multip. & Irreducible representation & Dim & Multip.\\
 				\hline
 				$\chi_0,\chi_0;e$ & 1 & 3 & $\chi_0,\chi_1;e$ & 2 &  2 \\
 				$\chi_1,\chi_1;e$ & 1 & 1 & $\chi_0,\chi_3;e$ & 2 &2\\
 				$\chi_3,\chi_3;e$ & 1 &  1 & $\chi_1,\chi_3;e$ & 2 & 1 \\
 				$\chi_0,\chi_0,\rm{sgn}$ & 1 & 1 & & &\\
 				\hline
 		\end{tabular}
		\caption{Two $T$-gates decomposition.}
	\end{table*}
 \begin{table*}
		%
		\begin{tabular}{|c|c|c||c|c|c||c|c|c|}
 				\hline
 				Irreducible representation & Dim & Multip. & Irreducible representation & Dim & Multip. & Irreducible representation & Dim & Multip.\\
 				\hline
 				$\chi_0,\chi_0,\chi_0;e$ & 1 & 4 & $\chi_0,\chi_0,\chi_3;e$ & 3 & 3 & $\chi_0,\chi_0,\chi_1;\rm{sgn}$ & 3 & 1 \\
 				$\chi_1,\chi_1,\chi_1;e$ & 1 & 1 & $\chi_0,\chi_1,\chi_1;e$ & 3 & 2 & $\chi_0,\chi_0,\chi_3;\rm{sgn}$ & 3 & 1 \\
 				$\chi_3,\chi_3,\chi_3;e$ & 1 & 1 &  $\chi_1,\chi_1,\chi_3;e$ & 3 & 1 & $\chi_0,\chi_1,\chi_3;e$ & 6 & 2 \\
 				$\chi_0,\chi_0,\chi_0;\,{\rm std}$ & 2 & 2 &  $\chi_0,\chi_3,\chi_3;e$ & 3 & 2 &&&\\
 				$\chi_0,\chi_0,\chi_1;e$ & 3 & 3 & $\chi_1,\chi_3,\chi_3;e$ & 3 & 1 &&&\\
 				\hline
 		\end{tabular}
		\caption{Three $T$-gates decomposition.}
		\end{table*}
 \begin{table*}	
	\begin{tabular}{|c|c|c||c|c|c||c|c|c|}
 				\hline
 				Irreducible representation & Dim & Multip. & Irreducible representation & Dim & Multip. & Irreducible representation & Dim & Multip.\\
 				\hline
 				$\chi_0,\chi_0,\chi_0,\chi_0;e$ & 1 & 5 & $\chi_1,\chi_1,\chi_1,\chi_3;e$ & 4 & 1 & $\chi_1,\chi_1,\chi_3,\chi_3;e$ & 6 & 1 \\
 				
 				$\chi_1,\chi_1,\chi_1,\chi_1;e$ & 1 & 1 & $\chi_0,\chi_3\chi_3,\chi_3;e$ & 4 & 2 & $\chi_0,\chi_0,\chi_1,\chi_1;{\rm ker}_a$ & 6 & 1 \\
 				
 				$\chi_3,\chi_3,\chi_3,\chi_3;e$ & 1 & 1 &  $\chi_1,\chi_3\chi_3,\chi_3;e$ & 4 & 1 & $\chi_0,\chi_0,\chi_3,\chi_3;{\rm ker}_a$ & 6 & 1 \\
 				
 				$\chi_0,\chi_0,\chi_0,\chi_0;2{\rm dim}$ & 2 & 1 &  $\chi_0,\chi_0,\chi_0,\chi_1;{\rm std}$ & 8 & 2 & $\chi_0,\chi_0,\chi_1,\chi_3;e$ & 12 & 3\\
 				
 				$\chi_0,\chi_0,\chi_0,\chi_0;{\rm std}$ & 3 & 3 & $\chi_0,\chi_0,\chi_0,\chi_3;{\rm std}$ & 8 & 2 & $\chi_0,\chi_1,\chi_1,\chi_3;e$ & 12 & 2\\
 				
 				$\chi_0,\chi_0,\chi_0,\chi_1;e$ & 4 & 4 & $\chi_0,\chi_0,\chi_1,\chi_1;e$ & 6 & 3 &
 				$\chi_0,\chi_1,\chi_3,\chi_3;e$ & 12 & 2\\
 				
 				$\chi_0,\chi_0,\chi_0,\chi_3;e$ & 4 & 4 & $\chi_0,\chi_0,\chi_3,\chi_3;e$ & 6 & 3 & $\chi_0,\chi_0,\chi_1,\chi_3;{\rm sgn}$ & 12 & 1\\
 				
 				$\chi_0,\chi_1,\chi_1,\chi_1;e$ & 4 & 2 &  &  &  &&&\\
 				\hline
 		\end{tabular}
		\caption{Four $T$-gates decomposition.}
		\end{table*}
We have obtained the irreducible decompositions for up to four tensor copies of the $T$ gate and report in 
{Tables I-IV}
the decomposition of each noise matrix. The superscripts of $\chi$ label the irreducible representations of $A_n$, while after the semicolon
we denote the irreducible representation of the stabilizer group, where $e$ denotes the trivial representation, ${\rm sgn}$ the sign representation, $std$ the standard representation for all subgroups of $S_4$, ${\rm ker}_a$ the \emph{Kernel $a$} representation of the Klein 4 group isomorphic to $S_2 \times S_2$, 
while $2{\rm dim}$ denotes the 2-dimensional representation of $S_4$.
We note that $\chi_2$ never appears in the decomposition, and that the highest multiplicity, being $n+1$, is always related to the trivial representation of the full group $G$.
Additionally, we note that exploiting Schur's Lemma and the above consideration, the number of $\lambda_j$ to be fitted when benchmarking copies of the $T$-gate is $\sum_{\alpha \, \mathrm{irrep}} m_\alpha$, i.e., from 1 to 4 qubits, this number is $4, 11, 24, 46$.


\section{{Classical recovery and estimation methods}}
In this section, we give details on the our numerical methods employed to {recover} the average fidelity from the measurement data provided by our protocol. In principle, we could try to obtain the parameters directly by a non-linear fitting approach along the lines of the variable projection algorithm that separates fitting of linear and non-linear fitting parameters \cite{OLeary2013}. However, given the fact, that the quantities we want to obtain correspond to the estimation of different decay rates in the data set, we are going to use instead a matrix pencil method for the extraction of signal poles developed in the context of 
\emph{signal processing}~\cite{MatrixPencilMethod,MatrixPencilMethodExp}. Due to the random nature of the sampling paths, this is supplemented by a bootstrapping approach in order to get reliable estimates on these parameters. We will continue describing these {two} 
components of our methodology in detail now, before commenting on the single and two $T$-gate examples described in the main text.

\subsection{Matrix pencil methods}\label{app:matrixPencilMeth}
{Key to the functioning of our scheme is the use of sophisticated methods of estimation which we lay out here.}
According to \eqref{eq:relation_fidelity_and_trace}, we can express the average fidelity $\EE({\mathcal F_{\Lambda^G, \mathcal I}})$ of the twirled channel in terms of the model parameters $\lambda_j$ as
\begin{equation}
\EE({\mathcal F_{\Lambda^G, \mathcal I}})
 =
 \frac{\sum \lambda_j +d}{d(d+1)} .
 \end{equation}
The relation between the model parameters $\lambda_j$ and the measurement data on the other hand is given by the zeroth-order fitting model
\begin{equation}\label{eq:zeroOrdModII}
F_{\mathrm{avg}}^{(0)}(\ell, \rho)
 =
 \sum_{j=1}^{4^n} (\lambda_j\, d_j)^\ell \xi_j\;.
 \end{equation}
From an abstract point of view this means that up to higher order terms, for each $\ell = 1,\dots, \ell_{\rm max}$, the measurement result $F_\ell$ can be expressed as a sum exponentially decaying terms of the form
\begin{equation}\label{eq:abstrSig}
  F_\ell = \sum_{j}^M \xi_j x_j^\ell\,
\end{equation}
 where we set $x_j=\lambda_j d_j$. Recovering the parameters $x_j$ from such a noisy data set is a well studied problem in the context of signal reconstruction going back to the work by Prony. Modern algorithms known as the \emph{Estimation of signal parameters via rotational (ESPRIT)} 
 rely on feature extraction via singular value decompositions and matrix pencils~\cite{MatrixPencilMethodDumped}. For a recent review and details of the construction see, e.g., Ref.~\cite{Prony} in the following, we give a short overview over the algorithm.

 In order to extract at most $M_{\rm max}$ signal poles $x_i$ from $\{F(\ell)\}_{\ell=0}^{L_{\rm max}-1}$, we first form the $2L_{\rm max}-M_{\rm max}\times M_{\rm max}+1$  \emph{Hankel matrix} $H_{\ell,k}= F(\ell+k)$ and compute its singular value decomposition 
 \begin{equation}
 H = U D V^*.
 \end{equation} 
 If there were no noise in the data, the number of non-zero singular values of $H$ would correspond directly to the number of unique {$\{x_i\}$} in \eqref{eq:abstrSig}. However, in the presence of noise, $H$ will typically 
{feature} full rank and we have to fix a threshold $\sigma_{\rm min}$ for the singular values to obtain a rank $M$ approximation of $H$ in order to extract $M\leq M_{\rm max}$ poles for the approximation of our measurement data. Let us denote by
 $W$ the $M\times M_{\rm max}+1$ matrix formed by the first $M$ rows of $V^*$ and set the $M_{\rm max}\times M$ matrices $V_0$ and $V_1$ equal to
 $(V_i)_{\ell,k} = M_{k,\ell+i}$, $\ell=1,\dots, M_{\rm max}$, $k=1,\dots {,} M$. The estimate for the signal poles $x_i$ can now be computed as the $M$ eigenvalues of the matrix $(V_0)^{-1}\cdot V_1$, where the inverse of $V_0$ is defined as its Moore-Penrose pseudo-inverse.
{In this way, we can extract from \eqref{eq:zeroOrdModII} the parameters $\{x_j\}=\{\lambda_j d_j\}$.} 

Now in principle, we would be left with the problem of matching the right $x_j$ with the correct $d_j$ in order to obtain the parameters $\lambda_j$. However, since typical quantum gates and in particular the $T$-gate have eigenvalues {which} are {close to being} roots of unity, we can {exploit} 
this property to circumvent this problem. Namely, assuming that we can find an $\tau$ such that $d_j^\tau=1$ for all $j$, we can partition our measurement data $\{F(\ell)\}_{\ell=1}^{\ell_{\rm max}}$ into $\tau$ subsets of the form $\{F(r), F(r+\tau), F(r+2\tau),\dots\}$ with $r=1,\dots, \tau$. In each of these subsets \eqref{eq:zeroOrdModII}
{reads} as
\begin{align}
  F_r (\ell) = \sum_j \xi_j d_j^r \left(d_j^\tau\right)^\ell \left(\lambda_j^\tau\right)^\ell= \sum_j \widetilde{\xi}_{j,r}  \left(\lambda_j^\tau\right)^\ell\;,
\end{align}
which is again of the form \eqref{eq:abstrSig} with $x_j=\lambda_j^\tau$ and $\widetilde{\xi}_{j,r} = \xi_j d_j^r$. Now using the reconstruction method described before, we can extract from this data $\lambda_j^\tau$ and in turn obtain $d_j$ by taking the $\tau$s-root. In terms of data collection from an experiment, we now have two options: we can restrict the collection of data points, i.e., measured expectation values, to $\ell=1,\tau, 2\tau,\dots$ or we can also collect data at intermediate points and extract the poles from the combined Hankel matrix $[H_1,H_2,\dots {,}H_\tau]$ in the way described before according to a multi-channel signal reconstruction approach~\cite{PoleEstimation}.

\subsection{Restriction to symmetry subspaces}

{In this subsection, we will explain how the restriction to symmetry subspaces will lead to a robust numerical recovery procedure.}
Under our standing assumption {that} the twirled noise channel $\Lambda^G_U$ is almost jointly diagonalizable with the target unitary channel $\mathcal{U}$, it follows directly, that  $\Lambda^G_U$ will also approximately preserve the invariant subspaces of the symmetry group $G$.
For our protocol this means that by choosing an appropriate initial states $\rho$ supported on such an invariant subspace, $(\Lambda^G_U)^\ell(\rho)$ will still be approximately located there. This observation serves us in two ways: On the one hand, we can get additional information about the performance of the benchmarked operation with respect to a specific subspace, i.e., we can identify the subspaces on which the errors occur. On the other hand, we reduce the number of signal parameters $\lambda_j$ we have to extract from our data considerably if we restrict the initial state to a particular subspace. This becomes  particularly important when we come close to an optimal implementation where the $\lambda_j$ become closer and closer to being degenerate.
To apply this approach we first construct the seven projectors linked to the {irreducible representations} in 
Table
{II}
and then take the corresponding eigenvectors as a basis of the invariant subspaces. Subsequently, we need to construct from these vectors a set of density operators such that we can address each irreducible subspace at least once and target only few of those subspaces in a single iteration. This choice is clearly not unique; for our numerics, we selected as input state the following density matrices. {It is a key insight to the functioning of the method that we can address single or a few irreducible subspaces in each iteration to arrive at a reliable and robust method that is able to reliably discriminate between close poles.}

\subsection{Bootstrapping for parameter estimation}

In this section, we give further details on the our numerical methods used to extract the average fidelity from the measurement data provided by our protocol. The starting point is the observation that the data obtained experimentally by executing our protocol contain more information than necessary for the model: Instead of having access to the observed average survival probabilities for a given execution length $\ell$ solely, our protocol actually provides this information at the level of each of the randomly chosen sample paths of length $\ell$ individually.
This insights motivates the use of standard methods in statistical estimation referred to as \emph{bootstrapping procedures} \cite{Statistics}  in order to reliably extract the average fidelity. Bootstrapping techniques refer to random sampling methods with replacement, designed in order to assign measures of accuracy to sample estimates.

For this, note that the sampling paths both with respect to the sequence length $\ell$ as well as with respect to the given realization of length $\ell$ are chosen independently. Hence, given the set of measured survival probabilities $\{F_{\mathbf{k}_{\ell,q}}\}$  were $1\leq \ell\leq \ell_{\max}$ denotes the circuit length and $\mathbf{k}_{\ell,q}$ the $q$-th randomly chosen sample sequence of symmetry gates for length $\ell$, we are going to resample this set of with respect to $\ell$. More precisely, in order to create a single bootstrap sample, we pick for each fixed $\ell$ a random subset with replacement of $m$ elements from $\{F_{\mathbf{k}_{\ell,q}}\}$ and compute their average with respect to $q$. For each of these resampled sequences, we compute the approximated average fidelity, according to our fitting model and matrix pencil methods described in Appendix~\ref{app:matrixPencilMeth}. The final estimate for the averaged fidelity is then obtained as average over all taken bootstrap samples.

\section{Numerical {recovery}}
In this section, we apply {the recovery} procedure detailed in Appendix~\ref{app:matrixPencilMeth} for two significant examples of practical relevance that are problematic to handle with the common approach to randomized benchmarking: A single $T$-gate and the tensor product of two $T$-gates.
\subsection{Single $T$-gate-case}
\label{sec:numerics:details}
\begin{figure}[t!]
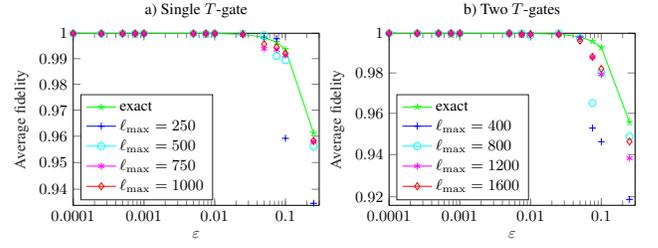

\includegraphics[width=0.23\textwidth]{SingleTgateDiffL.tikz}
\includegraphics[width=0.23\textwidth]{TwoTgateDiffL.tikz}\\

  \caption{Dependence of the estimated average fidelity for noisy single and two $T$-gate on the maximal sequence length   for different noise-strengths $\varepsilon$ extracted from bootstrapping over  $N=100$ sequences of length the given length (b). Green lines and stars indicate the analytic value of the average fidelity $\EE(\mc F)$ for the given noise level.}
 \label{fig:DifferLenghtTg2Tg}
\end{figure}

We describe here the numerics reported in the main text in greater detail. First, we consider the case of a single $T$-gate that is generated by the noisy Hamiltonian
\begin{align}\label{eq:noisTgatHam}
  H_\varepsilon = \frac{\pi}{8} \sigma_z - \varepsilon \, \sigma_x
\end{align}
with noise parameter {$\varepsilon>0$}. The four eigenvalues of the $T$-gate channel including multiplicities are given by ${\{1,1, {(1+\i)}/{\sqrt{2}},{(1-\i)}/{\sqrt{2}} \}}$. Accordingly, we have to extract four decay parameters $\lambda_j$ to estimate the average fidelity from \eqref{eq:zeroOrdModII}. As described in Appendix~\ref{app:matrixPencilMeth}, we make use of the fact that the $T^8=\1$ and hence we can extract the parameters $\lambda_j$ by  described matrix pencil approach. By exploiting the symmetry of the $T$-gate, we can obtain information in three invariant subspaces by running our protocol with the initial states
\begin{align}\label{app:eq:stateSingTgaNum}
  \rho\in\left\{\frac{1}{2}\begin{pmatrix}
                 1 & 0 \\
                 0 & 1
               \end{pmatrix}, \begin{pmatrix}
                 1 & 0\\
                 0&0
               \end{pmatrix}, \frac{1}{2}\begin{pmatrix}
                                1 & -\i \\
                                \i & 1
                              \end{pmatrix}\right\}
\end{align}
and measuring the corresponding survival probability. Fig.~\ref{fig:DifferLenghtTg2Tg} (a) shows the extracted estimates of the average fidelity depending on the strength of the perturbation $\varepsilon$ and the maximal sequence length of the protocol. It can be seen that we achieve a good agreement with the analytical value of the average fidelity even for short protocol sequences in the case of perturbations smaller than $\varepsilon\leq 0.01$. However, by increasing the sequence length to $\ell_{\max}=1000$, we even achieve satisfactory estimates starting from  $\varepsilon\leq 0.1$.

\subsection{Two $T$-gate case}
In the case of two $T$-gates applied in parallel to two qubits, we consider the perturbed Hamiltonian
\begin{align}\label{eq:noisTgatHam2Tg}
  H_\varepsilon = \frac{\pi}{8}\left(\sigma_z\otimes\1 + \1 \otimes\sigma_z - \varepsilon \, \sigma_x\otimes\sigma_x\right)\,
\end{align}
with the additional noise-term $\sigma_x\otimes\sigma_x$. In order to extract the decay parameters $\lambda_j$, we follow the approach of the single $T$-gate, the main difference being that we now choose the initial state among eight density matrices supported on different symmetry subspaces (they are reported below this paragraph). In each of these subspaces, we extract the corresponding decay parameters $\lambda_j$, which after bootstrapping leads us to an estimate of the desired average fidelity. Fig.~\ref{fig:DifferLenghtTg2Tg} (b) depicts theses estimates both in dependence of the perturbation strength $\varepsilon$ as well as depending on the maximally protocol length. Similar to the single $T$-gate case, we see that $\varepsilon\leq 0.01$ already $\ell_{\max}=800$ gives satisfactory results. However, in order to obtain meaningful lower bounds for larger values of $\varepsilon$, we have to double  $\ell_{\max}$ to $1600$.\\

We now provide the {quantum} states used as input for our numerics. Each of them combines the totally mixed state with some of the basis vectors obtained by the projectors onto the irreducible subspaces. To obtain the diagonal elements of $\Lambda$ connected to the trivial subspaces, we choose the states
\begin{equation*}
	\left(
	\begin{array}{cccc}
	\frac{1}{4} & 0 & 0 & 0 \\
	0 & \frac{1}{4} & 0 & 0 \\
	0 & 0 & \frac{1}{4} & 0 \\
	0 & 0 & 0 & \frac{1}{4} \\
	\end{array}
	\right),\left(
	\begin{array}{cccc}
	\frac{1}{2} & 0 & 0 & 0 \\
	0 & 0 & 0 & 0 \\
	0 & 0 & 0 & 0 \\
	0 & 0 & 0 & \frac{1}{2} \\
	\end{array}
	\right) \text{and} \left(
	\begin{array}{cccc}
	\frac{1}{2} & 0 & 0 & 0 \\
	0 & \frac{1}{4} & 0 & 0 \\
	0 & 0 & \frac{1}{4} & 0 \\
	0 & 0 & 0 & 0 \\
	\end{array}
	\right).
\end{equation*}
With the first one we can extract a single pole and with each of the other two states an additional parameter. With
\begin{equation*}
\left(
\begin{array}{cccc}
\frac{1}{4} & 0 & 0 & -\frac{1}{4} \\
0 & \frac{1}{4} & 0 & 0 \\
0 & 0 & \frac{1}{4} & 0 \\
-\frac{1}{4} & 0 & 0 & \frac{1}{4} \\
\end{array}
\right)
\end{equation*}
we retrieve {two} new parameters (in addition to the one corresponding to the trivial representation linked to the totally mixed state), one for irrep $\chi_1,\chi_1;e$ and one for $\chi_3,\chi_3;e$. The fifth state, 
\begin{equation*}
	\left(
	\begin{array}{cccc}
	\frac{1}{4} & 0 & 0 & 0 \\
	0 & \frac{1}{2} & 0 & 0 \\
	0 & 0 & 0 & 0 \\
	0 & 0 & 0 & \frac{1}{4} \\
	\end{array}
	\right),
\end{equation*}
is a combination of the totally mixed state and the one belonging to the irrep $\chi_0,\chi_0,\rm{sgn}$. With the first of the following two density matrices, 
\begin{equation*}
	\left(
	\begin{array}{cccc}
	\frac{1}{4} & -\frac{\i}{8} & -\frac{\i}{8} & 0 \\
	\frac{\i}{8} & \frac{1}{4} & 0 & \frac{\i}{8} \\
	\frac{\i}{8} & 0 & \frac{1}{4} & \frac{\i}{8} \\
	0 & -\frac{\i}{8} & -\frac{\i}{8} & \frac{1}{4} \\
	\end{array}
	\right),\left(
	\begin{array}{cccc}
	\frac{1}{4} & -\frac{\i}{8} & -\frac{\i}{8} & 0 \\
	\frac{\i}{8} & \frac{1}{4} & 0 & -\frac{\i}{8} \\
	\frac{\i}{8} & 0 & \frac{1}{4} & -\frac{\i}{8} \\
	0 & \frac{\i}{8} & \frac{\i}{8} & \frac{1}{4} \\
	\end{array}
	\right),
\end{equation*}
we obtain two poles for $\chi_0,\chi_1;e$ and two for $\chi_0,\chi_3;e$ (and again the trivial one). With the second operator with get the other two poles of these two {irreducible representations}.
The last density {matrix},
\begin{equation*}
	\left(
	\begin{array}{cccc}
	\frac{1}{4} & 0 & 0 & 0 \\
	0 & \frac{1}{4} & \frac{\frac{1}{4}-\frac{\i}{4}}{\sqrt{2}} & 0 \\
	0 & \frac{\frac{1}{4}+\frac{\i}{4}}{\sqrt{2}} & \frac{1}{4} & 0 \\
	0 & 0 & 0 & \frac{1}{4} \\
	\end{array}
	\right),
\end{equation*}
contains the basis vectors of the 2-dim irrep  $\chi_1,\chi_3;e$. With this selection, we cover every irreducible subspace and so we can extract all the $\lambda_j$ elements, addressing at most five of them at one time.


\begin{thebibliography}{10}

\bibitem{Roadmap}
A.~Acin, I.~Bloch, H.~Buhrman, T.~Calarco, C.~Eichler, J.~Eisert, D.~Esteve,
  N.~Gisin, S.~J. Glaser, F.~Jelezko, S.~Kuhr, M.~Lewenstein, M.~F. Riedel,
  P.~O. Schmidt, R.~Thew, A.~Wallraff, I.~Walmsley, and F.~K. Wilhelm.
\newblock {The European quantum technologies roadmap}.
\newblock {\em New J. Phys.}, 20:080201, 2018.

\bibitem{ChuangNielsenProcessTomography}
I.~L. Chuang and M.~A. Nielsen.
\newblock Prescription for experimental determination of the dynamics of a
  quantum black box.
\newblock {\em J. Mod. Opt.}, 44:2455--2467, 1997.

\bibitem{PhysRevA.77.032322}
M.~Mohseni, A.~T. Rezakhani, and D.~A. Lidar.
\newblock Quantum-process tomography: Resource analysis of different
  strategies.
\newblock {\em Phys. Rev. A}, 77:032322, 2008.

\bibitem{Compressed}
D.~Gross, Y.-K. Liu, S.~T. Flammia, S.~Becker, and J.~Eisert.
\newblock Quantum state tomography via compressed sensing.
\newblock {\em Phys. Rev. Lett.}, 105:150401, 2010.

\bibitem{AverageGateFidelities}
I.~Roth, R.~Kueng, S.~Kimmel, Y.-K. Liu, D.~Gross, J.~Eisert, and M.~Kliesch.
\newblock Recovering quantum gates from few average gate fidelities.
\newblock {\em Phys. Rev. Lett.}, 121:170502, 2018.

\bibitem{Diamond}
M.~Kliesch, R.~Kueng, J.~Eisert, and D.~Gross.
\newblock Improving compressed sensing with the diamond norm.
\newblock {\em IEEE Trans. Inf. Th.}, 62:7445, 2016.

\bibitem{QuantumFieldTomography}
A.~Steffens, M.~Friesdorf, T.~Langen, B.~Rauer, T.~Schweigler, R.~H{\"u}bener,
  J.~Schmiedmayer, C.~A. Riofrio, and J.~Eisert.
\newblock Towards experimental quantum field tomography with ultracold atoms.
\newblock {\em Nature Comm.}, 6:7663, 2015.

\bibitem{MPSTomographyIons}
B.~P. Lanyon, C.~Maier, M.~Holz{\"a}pfel, T.~Baumgratz, C.~Hempel, P.~Jurcevic,
  I.~Dhand, A.~S. Buyskikh, A.~J. Daley, M.~Cramer, M.~B. Plenio, R.~Blatt, and
  C.~F. Roos.
\newblock Efficient tomography of a quantum many-body system.
\newblock arXiv:1612.08000.

\bibitem{PreskillFaultTolerant}
J.~Preskill.
\newblock Fault-tolerant quantum computation.
\newblock {\em arXiv:quant-ph/9712048}, 1997.

\bibitem{MartinisFowlerFaultTolerant}
R.~Barends, J.~Kelly, A.~Megrant, A.~Veitia, D.~Sank, E.~Jeffrey, T.~C. White,
  J.~Mutus, A.~G. Fowler, B.~Campbell, Y.~Chen, Z.~Chen, B.~Chiaro,
  A.~Dunsworth, C.~Neill, P.~O'Malley, P.~Roushan, A.~Vainsencher, J.~Wenner,
  A.~N. Korotkov, A.~N. Cleland, and J.~M. Martinis.
\newblock Superconducting quantum circuits at the surface code threshold for
  fault tolerance.
\newblock {\em Nature}, 508, 2014.

\bibitem{IBMFaultTolerant}
J.~M. Chow, J.~M. Gambetta, E.~Magesan, S.~J. Srinivasan, A.~W. Cross, D.~W.
  Abraham, N.~A. Masluk, B.~R. Johnson, C.~A. Ryan, and M.~Steffen.
\newblock Implementing a strand of a scalable fault-tolerant quantum computing
  fabric.
\newblock {\em Nature Comm.}, 5:4015, 2014.

\bibitem{FirstRB}
J.~Emerson, R.~Alicki, and K.~Zyczkowski.
\newblock Scalable noise estimation with random unitary operators.
\newblock {\em J. Opt. B}, 7:S347--S352, 2005.

\bibitem{PhysRevA.80.012304}
C.~Dankert, R.~Cleve, J.~Emerson, and E.~Livine.
\newblock Exact and approximate unitary 2-designs and their application to
  fidelity estimation.
\newblock {\em arXiv:0606161}, 2006.

\bibitem{KnillBenchmarking}
E.~Knill, D.~Leibfried, R.~Reichle, J.~Britton, R.~B. Blakestad, J.~D. Jost,
  C.~Langer, R.~Ozeri, S.~Seidelin, and D.~J. Wineland.
\newblock Randomized benchmarking of quantum gates.
\newblock {\em Phys. Rev. A}, 77:012307, 2008.

\bibitem{MagGamEmer}
E.~Magesan, J.~M. Gambetta, and J.~Emerson.
\newblock Characterizing quantum gates via randomized benchmarking.
\newblock {\em Phys. Rev. A}, 85:042311, 2012.

\bibitem{MagGamEmer2}
E.~Magesan, J.~M. Gambetta, and J.~Emerson.
\newblock Robust randomized benchmarking of quantum processes.
\newblock {\em Phys. Rev. Lett.}, 106:042311, 2011.

\bibitem{RBFlammia}
J.~J. Wallman and S.~T. Flammia.
\newblock Randomized benchmarking with confidence.
\newblock {\em New. J. Phys.}, 16:103032, 2014.

\bibitem{EmersonDiscussion}
K.~Boone, A.~Carignan-Dugas, J.~J. Wallman, and J.~Emerson.
\newblock Randomized benchmarking under different gatesets.
\newblock {\em arXiv:1811.01920}.

\bibitem{GroAudEis}
D.~Gross, K.~Audenaert, and J.~Eisert.
\newblock Evenly distributed unitaries: on the structure of unitary designs.
\newblock {\em J. Math. Phys.}, 48:052104, 2007.

\bibitem{Cross16}
A.~W. Cross, E.~Magesan, L.~S. Bishop, J.~A. Smolin, and J.~M. Gambetta.
\newblock Scalable randomised benchmarking of non-{C}lifford gates.
\newblock {\em npj Quant. Inf.}, 2, 2016.

\bibitem{DugWallEme15}
C.~A. Dugas, J.~Wallman, and J.~Emerson.
\newblock Characterizing universal gate sets via dihedral benchmarking.
\newblock {\em Phys. Rev. A}, 92:060302, 2015.

\bibitem{InterleavedRB}
E.~Magesan, J.~M. Gambetta, B.~R. Johnson, C.~A. Ryan, J.~M. Chow, S.~T.
  Merkel, M.~P. da~Silva, G.~A. Keefe, M.~B. Rothwell, T.~A. Ohki, M.~B.
  Ketchen, and M.~Steffen.
\newblock Efficient measurement of quantum gate error by interleaved randomized
  benchmarking.
\newblock {\em Phys. Rev. Lett.}, 109:080505, 2012.

\bibitem{Harper}
R.~Harper and S.~T Flammia.
\newblock {Estimating the fidelity of T gates using standard interleaved
  randomized benchmarking}.
\newblock {\em Quant. Sc. Tech.}, 2:015008, 2017.

\bibitem{IndependentNoise}
J.~J. Wallman.
\newblock Randomized benchmarking with gate-dependent noise.
\newblock {\em Quantum}, 2:47, 2018.

\bibitem{RBFiniteGroups}
D.~S. Franca and A.~K. Hashagen.
\newblock Approximate randomized benchmarking for finite groups.
\newblock {\em J. Phys. A}, 51:395302, 2018.

\bibitem{WintonRB}
W.~G. Brown and B.~Eastin.
\newblock Randomized benchmarking with restricted gate sets.
\newblock {\em Phys. Rev. A}, 97:062323, 2018.

\bibitem{MatrixPencilMethodDumped}
Y.~Hua and T.~K. Sarkar.
\newblock Matrix pencil method for estimating parameters of exponentially
  damped/undamped sinusoids in noise.
\newblock {\em IEEE Transactions on Acoustics, Speech, and Signal Processing},
  38:814--824, May 1990.

\bibitem{KimmOhki}
S.~Kimmel, M.~P. da~Silva, C.~A. Ryan, B.~R. Johnson, and T.~Ohki.
\newblock Robust extraction of tomographic information via randomized
  benchmarking.
\newblock {\em Phys. Rev. X}, 4:011050, 2014.

\bibitem{GlasBron}
K.~Glashoff and M.~M. Bronstein.
\newblock Almost-commuting matrices are almost jointly diagonalizable, 2013.
\newblock arXiv:1305.2135.

\bibitem{AndSosaRioDeuJes}
B.~E. Anderson, H.~Sosa-Martinez, C.~A. Riofrio, I.~H. Deutsch, and P.~S.
  Jessen.
\newblock Accurate and robust unitary transformations of a high-dimensional
  quantum system.
\newblock {\em Phys. Rev. Lett.}, 114:240401, 2015.

\bibitem{Helsen17}
J.~Helsen, J.~Wallman, S.~T. Flammia, and S.~Wehner.
\newblock Multi-qubit randomized benchmarking using few samples, 2017.
\newblock arXiv:1701.04299.

\bibitem{Zassenhaus}
F.~Casas, A.~Murua, and M.~Nadinic.
\newblock {Efficient computation of the Zassenhaus formula}.
\newblock {\em Comp. Phys. Comm.}, 183:2386--2391, 2012.

\bibitem{Serre}
J.-P. Serre.
\newblock {\em Linear representations of finite groups}.
\newblock Springer, 1996.

\bibitem{Berndt}
R.~Berndt.
\newblock {\em Representations of linear groups}.
\newblock Vieweg, first edition edition, 2007.

\bibitem{Nielsen02}
M.~A. Nielsen.
\newblock A simple formula for the average gate fidelity of a quantum dynamical
  operation.
\newblock {\em Phys. Rev. A}, 303:249--252, 2002.

\bibitem{OLeary2013}
D.~P. O'Leary and B.~W. Rust.
\newblock Variable projection for nonlinear least squares problems.
\newblock {\em Computational Optimization and Applications}, 54:579--593, 2013.

\bibitem{MatrixPencilMethod}
Y.~Hua and T.~K. Sarkar.
\newblock Matrix pencil method for estimating parameters of exponentially
  damped/undamped sinusoids in noise.
\newblock {\em IEEE Trans. Ac. Sp. Sig. Proc.}, 38:814--824, May 1990.

\bibitem{MatrixPencilMethodExp}
T.~K. Sarkar and O.~Pereira.
\newblock Using the matrix pencil method to estimate the parameters of a sum of
  complex exponentials.
\newblock {\em IEEE Ant. Prop. Mag.}, 37:48--55, Feb 1995.

\bibitem{Prony}
D.~Potts and M.~Tasche.
\newblock Parameter estimation for nonincreasing exponential sums by prony-like
  methods.
\newblock {\em Linear Algebra and its Applications}, 439:1024 -- 1039, 2013.
\newblock 17th Conference of the International Linear Algebra Society,
  Braunschweig, Germany, August 2011.

\bibitem{PoleEstimation}
J.-M. Papy, L.~De Lathauwer, and S.~Van Huffel.
\newblock Common pole estimation in multi-channel exponential data modeling.
\newblock {\em Signal Processing}, 86:846 -- 858, 2006.

\bibitem{Statistics}
D.~S. Moore, G.~P. McCabe, and B.~A. Craig.
\newblock {\em Introduction to the practice of statistics}.
\newblock W. H. Freeman, 2016.

\end{thebibliography}
\end{document}